\documentclass[reprint,amsmath, superscriptaddress, amssymb, aps, prb]{revtex4-2}
\pdfoutput=1
\usepackage{lipsum}  
\usepackage{stackengine}
\usepackage{graphicx}
\usepackage{amsmath}
\usepackage{soul}
\usepackage{dcolumn}
\usepackage{comment}
\usepackage{bm}
\usepackage{color}
\usepackage[normalem]{ulem}
\begin{document}
\renewcommand{\thesection}{\Roman{section}}
\preprint{APS/123-QED}
\title{Electrically tunable dipolar interactions between layer-hybridized excitons}
\author{Daniel Erkensten}
\affiliation{Department of Physics, Chalmers University of Technology, 41296 Gothenburg, Sweden}
\author{Samuel Brem}
\affiliation{Department of Physics, Philipps-Universit{\"a}t Marburg, 35037 Marburg, Germany}
\author{Ra\"ul Perea-Caus\'in}
\affiliation{Department of Physics, Chalmers University of Technology, 41296 Gothenburg, Sweden}
\author{Joakim Hagel}
\affiliation{Department of Physics, Chalmers University of Technology, 41296 Gothenburg, Sweden}
\author{Fedele Tagarelli}
\affiliation{Institute of Electrical and Microengineering, École Polytechnique Fédérale de Lausanne (EPFL), Lausanne, Switzerland}
\author{Edoardo Lopriore}
\affiliation{Institute of Electrical and Microengineering, École Polytechnique Fédérale de Lausanne (EPFL), Lausanne, Switzerland}
\author{Andras Kis}
\affiliation{Institute of Electrical and Microengineering, École Polytechnique Fédérale de Lausanne (EPFL), Lausanne, Switzerland}
\author{Ermin Malic}
\affiliation{Department of Physics, Philipps-Universit{\"a}t Marburg, 35037 Marburg, Germany}
\affiliation{Department of Physics, Chalmers University of Technology, 41296 Gothenburg, Sweden}
\begin{abstract}
Transition-metal dichalcogenide bilayers exhibit a rich exciton landscape including layer-hybridized excitons, i.e. excitons which are of partly intra- and interlayer nature. In this work, we study hybrid exciton-exciton interactions in naturally stacked WSe$_2$ homobilayers. In these materials, the exciton landscape is electrically tunable such that the low-energy states can be rendered more or less interlayer-like depending on the strength of the external electric field.
Based on a microscopic and material-specific many-particle theory, we reveal two intriguing interaction regimes: a low-dipole regime at small electric fields and a high-dipole regime at larger fields, involving interactions between hybrid excitons with a substantially different intra- and interlayer composition in the two regimes. While the low-dipole regime is characterized by weak inter-excitonic interactions between intralayer-like excitons, the high-dipole regime involves mostly interlayer-like excitons which display a strong dipole-dipole repulsion and give rise to large spectral blue-shifts and a highly anomalous diffusion. Overall, our microscopic study sheds light on the remarkable electrical tunability of hybrid exciton-exciton interactions in atomically thin semiconductors and can guide future experimental studies in this growing field of research.
\end{abstract}
\maketitle

Recently, two-dimensional van der Waals heterostructures, formed by stacking transition-metal dichalcogenide monolayers on top of each other, have emerged as a promising platform for engineering strong correlations, topology and intriguing many-body interactions \cite{geim2013van, mak2022semiconductor, xu2022tunable, mueller2018exciton, perea2022exciton}. In particular, these structures exhibit spatially separated interlayer excitons, i.e., Coulomb-bound electron-hole pairs where the constituent electrons and holes reside in different layers, which display permanent out-of-plane dipole moments \cite{nagler2017interlayer, merkl2019ultrafast, schmitt2022formation, miller2017long, ciarrocchi2022excitonic}. Furthermore, intra- and interlayer exciton states can be efficiently hybridized via electron and hole tunneling, and form new hybrid excitons (hX) that inherit properties of both exciton species \cite{alexeev2019resonantly, PhysRevB.99.125424, brem2020hybridized, peimyoo2021electrical}. The formation of hybrid excitons is particularly favorable in naturally stacked homobilayers, as opposed to type-II heterostructures where the dominating exciton species have mostly interlayer character \cite{PhysRevB.97.165306, latini2017interlayer}. 

\begin{figure}[t!]
    \centering
    \includegraphics[width=\columnwidth]{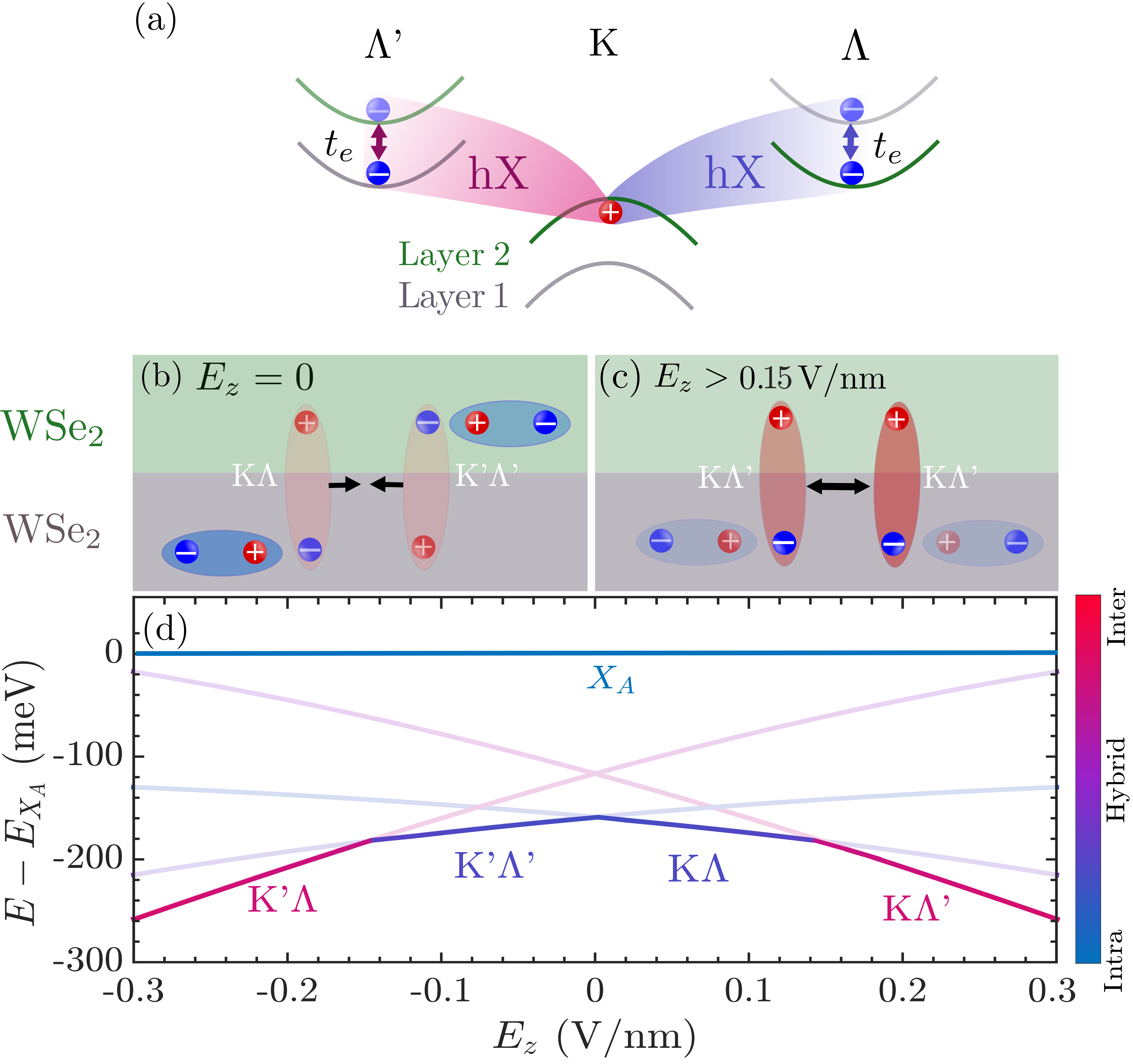}
    \caption{Hybrid exciton species in naturally stacked bilayer WSe$_2$. \textbf{(a):} K$\Lambda$ and K$\Lambda$' hybrid excitons (hX) are primarily formed from electrons and holes in the same or different layers, respectively. The two layers are indicated with green and gray lines. The hybridization is predominantly induced via electron ($t_e$) tunneling. Note that the K'$\Lambda$' and K'$\Lambda$ excitons, which are degenerate with the respective K$\Lambda$ and K$\Lambda$' states and exhibit opposite dipole moments, are not shown. \textbf{(b):} At vanishing electric fields the degenerate K$\Lambda$ and K'$\Lambda$' hX are the energetically lowest states. These are mostly intralayer-like in nature and attract each other due to their opposite dipole moments. \textbf{(c):} At electric fields $E_z>0.15$ V/nm, K$\Lambda$' excitons constitute the energetically lowest states. These are mostly interlayer in nature and repel each other. \textbf{(d):} Exciton landscape as a function of electric field, with the colorbar revealing a transition from a mostly intralayer-like  (blue) to a mostly interlayer-like exciton state (red) at elevated electric fields. The thick line denotes the energetically lowest state that changes depending on the electric field. The energies are given with respect to the KK intralayer exciton (usually denoted as $X_A$  in literature).  }
    \label{schematicfig}
\end{figure}

Moreover, the ground state of hybrid excitons can be optically inactive or momentum-dark \cite{kunstmann2018momentum}, as is the case in WSe$_2$ homobilayers (Fig. \ref{schematicfig}(a)) \cite{deilmann2019finite, hagel2021exciton}. Here, the
efficient electron tunneling at the $\Lambda$-point of the Brillouin zone (and less efficient hole tunneling at the K-point) results in a strongly hybridized K$\Lambda$ exciton state \cite{brem2020hybridized}. Furthermore, in naturally stacked  H-type (2H) WSe$_2$ homobilayers, the K$\Lambda$ state is energetically degenerate with the K'$\Lambda$' state, however these two states exhibit opposite dipole orientations (as a consequence of the inverted spin-orbit splitting in one of the layers \cite{brem2020hybridized}). This stacking configuration also enables the formation of other exciton species, such as K$\Lambda$' excitons, which lie energetically close to the degenerate K$\Lambda$ and K'$\Lambda$' states. Intriguingly, the K$\Lambda$' exciton state exhibits a much larger interlayer component than the K$\Lambda$ state as schematically illustrated in Fig. \ref{schematicfig}(a) (where green and gray bands refer to the upper and lower TMD layer, respectively). As hybrid excitons are partly of interlayer character, they also exhibit an out-of-plane dipole moment which couples to externally applied electric fields via the quantum-confined Stark effect \cite{leisgang2020giant, jauregui2019electrical, wang2018electrical, lopriore2022ultrafast, deilmann2018interlayer}, such that the interlayer component of these excitons and even the ordering of different hybrid exciton states can be tuned \cite{hagel2022electrical, PhysRevB.105.L041409}. This implies that also the interactions, in particular the dipole-dipole repulsion, between different types of hybrid excitons should be electrically tunable. Hence, a remarkable number of fundamentally and technologically relevant phenomena governed by exciton-exciton interactions in TMDs could potentially be electrically controlled. Some of these phenomena include experimentally observed blue-shifts of exciton resonances with excitation power \cite{nagler2017interlayer, unuchek2019valley}, anomalous exciton transport \cite{yuan2020twist, sun2021excitonic}, and even the stability of Bose-Einstein condensates \cite{wang2019evidence, shi2022bilayer}, the conditions for superfluidity \cite{gotting2022moire, PhysRevB.103.L041406, fogler2014high} and the exciton compressibility that is important for the characterization of excitonic insulators \cite{ma2021strongly}.

In this work, we develop a material-specific and predictive many-particle theory of hybrid exciton-exciton interactions using the density matrix formalism. We investigate the impact of electric fields on density-dependent energy renormalizations and exciton transport at elevated excitation densities in naturally stacked WSe$_2$ homobilayers. We show that two intriguing interaction regimes emerge when applying an out-of-plane electric field: \textbf{(i)} a low-dipole regime at  $E_z\lesssim$ 0.15 V/nm (Fig. \ref{schematicfig}(b)), where interactions are governed by mostly intralayer-like K$\Lambda$ and K'$\Lambda$' excitons which mutually attract each other, and \textbf{(ii)} a high-dipole regime at $E_z\gtrsim $ 0.15 V/nm (Fig. \ref{schematicfig}(c)), where mostly interlayer-like K$\Lambda$' excitons constitute the energetically lowest state which exhibits a strong dipole-dipole repulsion. These regimes give rise to substantially different behaviors for the experimentally accessible energy renormalizations and exciton transport. While the low-dipole regime is characterized by negligible exciton line-shifts and a conventional diffusion,  the high-dipole regime exhibits considerable blue-shifts of tens of meVs and a highly anomalous diffusion. Overall, our work provides a recipe for future experiments on how to tune the hybrid exciton-exciton interaction and in particular exciton transport at elevated excitation powers.    

\section*{Hybrid exciton landscape}
To model exciton-exciton interactions between layer-hybridized excitons in TMD bilayers, we first set up an excitonic Hamilton operator $H=H_{x,0}+H_{x-x}$ expressed in a monolayer eigenbasis \cite{katsch2018theory}. Here, the first part of the Hamiltonian takes into account the centre-of-mass motion of intra- and interlayer excitons, their Coulomb binding, and their hybridization via an effective tunneling model. The Hamiltonian which captures exciton hybridization reads \cite{brem2020hybridized, merkl2020twist}
\begin{align}
\begin{split}
H_{x,0}=\sum_{\xi, L,L', \bm{Q}}(E^{\xi}_{L, \bm{Q}}\delta_{L,L'}+T^{\xi}_{LL'})X^{\dagger \xi }_{L, \bm{Q}}X^{\xi}_{L', \bm{Q}} \ , 
\end{split}
\label{part1}
\end{align}
with the first term being the exciton dispersion $E^{\xi}_{L, \bm{Q}}=\frac{\hbar^2 \bm{Q}^2}{2M^{\xi L}}+E^{\xi L}_b+\Delta^{\xi L}$, and the exciton binding energy, $E^{\xi L}_b$, obtained from solving the bilayer Wannier equation \cite{kira2006many, hagel2021exciton}. Here, $M^{\xi L}$ is the total exciton mass, $L=(l_h, l_e)$ is a compound layer index, $\xi=(\xi_h, \xi_e)$ is the exciton valley and $\bm{Q}$ is the centre-of-mass momentum. Furthermore, $\Delta^{\xi L}$ contains the valley-specific band gap. Note that, due to the degeneracy between exciton states with different spin-valley configurations (neglecting electron-hole exchange \cite{PhysRevLett.115.176801}), it is sufficient to consider a single spin system, e.g. exciton states being formed by spin-up valence and conduction bands, so that spin indices can be omitted. The excitonic operators $X^{\dagger \xi}_{L, \bm{Q}}$ ($X^{\xi}_{L, \bm{Q}}$) create (annihilate) intralayer ($X$, $l_e=l_h$) or interlayer ($IX$, $l_e\neq l_h$) excitons. The second part of Eq. \eqref{part1} takes into account the tunneling of electrons and holes between different layers via the excitonic tunneling matrix element, $T^{\xi}_{LL'}$ \cite{hagel2021exciton}. The latter is dependent on electron/hole tunneling strengths and excitonic wave function overlaps, cf. Supplementary Section I for details. By performing the basis transformation $X^{\dagger \xi}_{L, \bm{Q}}=\sum_{\eta } Y^{\dagger \xi}_{\eta, \bm{Q}}C^{\xi \eta}_{L, \bm{Q}}$, introducing the \emph{hybrid} exciton operators $Y^{(\dagger)}_{\eta}$, the hybrid exciton state $\eta$, and the mixing coefficients $C$ determining the relative intra- and interlayer content of the hybrid exciton, the Hamiltonian in Eq. \eqref{part1} is diagonalized and becomes 
\begin{equation}
    \tilde{H}_{x,0}=\sum_{\xi, \eta, \bm{Q}} E^{(hX)\xi}_{\eta, \bm{Q}}Y^{\dagger \xi}_{\eta, \bm{Q}}Y^{\xi}_{\eta, \bm{Q}} \ , 
\end{equation}
where the hybrid exciton dispersion $E^{(hX)\xi}_{\eta, \bm{Q}}$ along with the mixing coefficients are obtained from solving the hybrid exciton eigenvalue problem, cf. Supplementary Section I. In this work, we are only concerned with the lowest hybrid exciton state $\eta$, and omit this index in the following. In Fig. \ref{schematicfig}(d), we show the hybrid exciton landscape for an hBN-encapsulated and naturally stacked (2H) WSe$_2$ homobilayer, including the four lowest-lying states. In Table S1 in Supplementary Section I, we include also higher-lying transitions. The hybrid exciton eigenenergy $E\equiv E^{(hX)}_{\bm{Q}=\bm{0}}$ is given relative to the intralayer $A$ exciton energy, $E_{X_A}$. We find that $\xi=(\xi_h, \xi_e)=\text{K}\Lambda$/K'$\Lambda$' exciton states constitute the energetically lowest states, lying approximately 160 meV below the bright $X_A$ state. The colorbar indicates the corresponding interlayer component of the mixing coefficient, revealing the hybrid nature of K$\Lambda$/K'$\Lambda$' excitons, cf. Fig. \ref{schematicfig}(d) at $E_z=0$. 

Furthermore, we study the exciton landscape as a function of an out-of-plane electric field, $E_z$. This is done by exploiting the electrostatic Stark shifts of the interlayer exciton energies, which influence the intra- and interlayer composition of hybrid excitons \cite{leisgang2020giant, wang2018electrical, deilmann2018interlayer}. Intriguingly, we find that for positive (negative) electric fields $|E_z|>0.15$ V/nm, the energetically lowest state corresponds to the K$\Lambda$' (K'$\Lambda$) state, i.e. the ordering of different exciton states is changed. 
This is explained by the fact that, as a consequence of the band-ordering (Fig. \ref{schematicfig}(a)), the K$\Lambda$' state carries a significantly larger interlayer component than the K$\Lambda$ state and is as such easier modulated with respect to electric fields. In particular, the K$\Lambda$' and K$\Lambda$ states possess an interlayer component $|C_{IX}|^2$ of 0.64 (0.80) and 0.23 (0.39), respectively, at $E_z=0 \ (0.3)$ V/nm. The fact that the dominating exciton species at elevated electric fields carries a large interlayer component and consequently a large dipole moment is also reflected in the stronger exciton-exciton interaction, as we shall see in the following.

\section*{Hybrid exciton-exciton interactions}
Now, we consider the interacting part of the Hamiltonian, $H_{x-x}$. In this work, we focus on the direct part of the interaction, and the contributions from interlayer excitons. Interlayer exchange interactions are seen to give a minor correction to the direct dipole-dipole interaction \cite{PhysRevB.92.165121, erkensten2022microscopic}.  Although intralayer exchange interactions (taking into account exchange of individual carriers) are dominant in TMD monolayers \cite{PhysRevB.96.115409, erkensten2021exciton, PhysRevB.96.115409}, they are known to have a negligible impact on experimentally accessible density-dependent energy renormalizations, as their contributions are largely cancelled out against contributions due to higher-order correlation effects \cite{trovatello2022disentangling,katsch2019theory}. This is also supported by recent experiments, which report negligible shifts with excitation power of intralayer exciton resonances and sizable blue-shifts in luminescence spectra for interlayer excitons \cite{nagler2017interlayer, yuan2020twist}. Furthermore, we assumed that the excitons can be treated as independent bosons, which holds in the weakly interacting limit $n_x a_B^2\ll1$, where $a_B$ is the exciton Bohr radius and $n_x$ is the exciton density \cite{de2001exciton}.

We transform the interaction Hamiltonian to the hybrid basis (cf. Supplementary Section II for details) resulting in 
\begin{equation}
    \tilde{H}_{x-x}=\frac{1}{2A}\sum_{\xi, \xi', \bm{q}, \bm{Q}, \bm{Q}'}\tilde{W}^{\xi\xi'}_{\bm{q}}Y^{\dagger \xi}_{\bm{Q}+\bm{q}}Y^{\dagger \xi'}_{\bm{Q}'-\bm{q}}Y^{\xi'}_{\bm{Q}'}Y^{\xi}_{\bm{Q}} \ ,
    \label{eqham}
\end{equation}
with the hybrid dipole-dipole interaction matrix element 
 $   \tilde{W}^{\xi\xi'}_{\bm{q}}=\sum_{i,j=1,2}W^{\xi\xi'}_{IX_i, IX_j, \bm{q}}|C^{\xi}_{IX_i}|^2|C^{\xi'}_{IX_j}|^2$
 and the normalization area $A$. The hybrid exciton-exciton interaction crucially includes the pure interlayer dipole-dipole interaction between different interlayer exciton species $IX_i$, $i=1,2$, weighted by the corresponding mixing coefficients. The interlayer dipole-dipole matrix element reads in the long wavelength limit $W^{\xi\xi'}_{IX_i, IX_i, \bm{q}\rightarrow \mathbf{0}}=-W^{\xi\xi'}_{IX_i, IX_j, \bm{q}\rightarrow \mathbf{0}}=\frac{d_{\mathrm{TMD}}e_0^2}{2\epsilon_0 \epsilon^{\perp}_{\mathrm{TMD}}}$, where $i\neq j$ and with $d_{\mathrm{TMD}}$ and $\epsilon^{\perp}_{\mathrm{TMD}}$ being the TMD thickness and the out-of-plane component of the dielectric tensor of the TMD, respectively. The sign difference between the interactions is a consequence of the opposite dipole orientations of the interlayer excitons $IX_1$ and $IX_2$ (cf. Fig. \ref{schematicfig}(b)). 
 The full hybrid Hamiltonian including intra- and interlayer direct and exchange contributions is derived in Supplementary Section II. 
\begin{figure}[t!]
    \centering
    \includegraphics[width=\columnwidth]{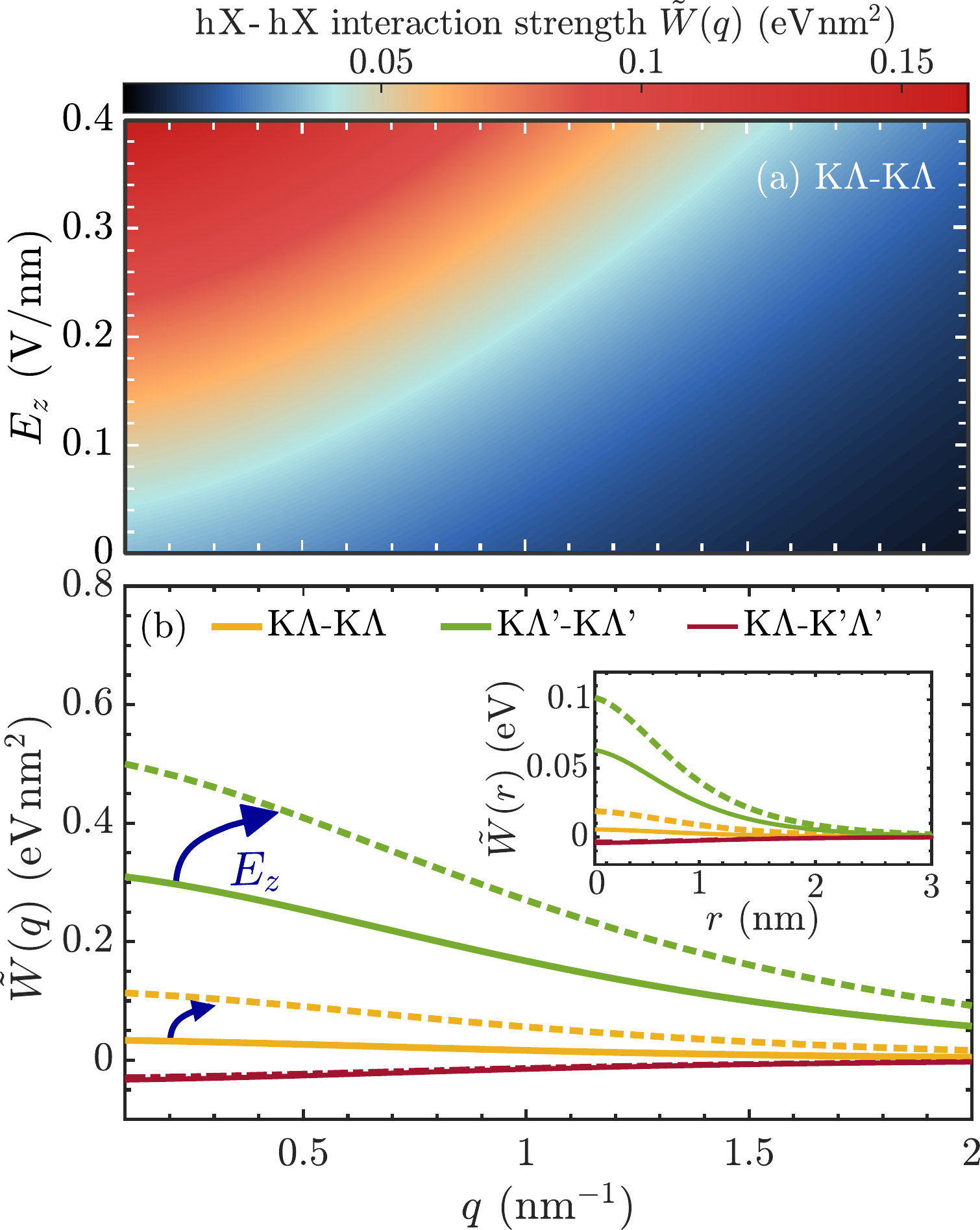}
    \caption{Hybrid exciton-exciton interactions. \textbf{(a):} Momentum- and electric-field dependent interaction strength involving K$\Lambda$ hX. The interaction strength increases as the interlayer component is enhanced under the application of an out-of-plane electric field, $E_z$. \textbf{(b):} Momentum-dependent interaction strengths for different electric fields and between different types of hybrid excitons. Solid lines correspond to the case $E_z=0$ and dashed lines to the case of $E_z=0.3$ V/nm. The inset shows the corresponding real-space interaction, $\tilde{W}(r)$.}
    \label{matrixelement}
\end{figure}

In Fig. \ref{matrixelement}(a), we display the hybrid exciton-exciton interaction matrix element for $\xi=\xi'=\text{K}\Lambda$ hybrid excitons as a function of momentum and out-of-plane electric field. The interaction is repulsive ($>0$) and maximized in the long wavelength limit. The interaction strength is also found to be highly tunable with respect to electric fields via its quartic dependence on interlayer mixing coefficients (cf. Eq. \eqref{eqham}). In particular, the interlayer component of hybrid excitons is enhanced with $E_z$, if the electric field is applied parallel to the dipole moment of the hX, cf. Fig. \ref{schematicfig}(c). In Fig. \ref{matrixelement}(b), we consider interactions between different types of hybrid exciton species at vanishing electric fields (solid lines) and at $E_z=0.3$ V/nm (dashed lines). Due to their large interlayer component (Fig. \ref{schematicfig}(d)), K$\Lambda$' excitons exhibit the strongest dipole-dipole repulsion, which is further enhanced with $E_z$. Furthermore, we note that the interaction between K$\Lambda$ and K'$\Lambda$' excitons is attractive ($<$0), cf. Fig. \ref{schematicfig}(b). This is a consequence of the interlayer component of these excitons having opposite dipole moments. As such, the exciton states energetically shift in opposite directions under the application of an electric field and the increase of interlayer component in one of the excitons is compensated by a decrease of interlayer component in the other exciton (Fig. \ref{schematicfig}(d)). This yields an interaction strength which is largely independent on electric field. We also show the hybrid exciton-exciton interaction in real space (inset in Fig. \ref{matrixelement}(b)), and identify the dipole-dipole-like character of the interaction between excitons of the same valley species at large distances, i.e. $\tilde{W}(r)\sim d^2_{\mathrm{TMD}}/r^3$ (cf. Supplementary Section III) \cite{PhysRevB.106.L081412, schindler2008analysis}. Finally, we note that the real-space exciton-exciton interaction, crucially including the dipole-dipole interaction, is a key ingredient in the Bose-Hubbard model, which can be exploited to investigate the conditions for different quantum phases of excitonic systems, such as superfluidity, in semiconductor moir{\'e} materials \cite{PhysRevB.103.L041406, gotting2022moire,lagoin2022extended}.  

Having microscopic access to the hybrid exciton-exciton interaction matrix elements enables us to study density-dependent energy renormalizations observable in photoluminescence (PL) spectra. In particular, given the large electrical tunability of exciton-exciton interactions, we expect that applying electric fields in combination with increasing pump power can be used to engineer substantial blue-shifts of exciton luminescence peaks. This offers an intriguing way of realizing strong many-body interactions in atomically thin semiconductors. Note that the relevant excitons in homobilayer WSe$_2$ are momentum-dark (Fig. \ref{schematicfig}(a)), and become only visible via phonon sidebands in low-temperature PL \cite{lindlau2018role, brem2020phonon, PhysRevB.105.L041409, wang2018electrical, he2020valley}. In our theoretical model, we derive the density-dependent energy renormalization $\delta E^{\xi}$ by evaluating the Heisenberg equation of motion for the hybrid polarization on a Hartree-Fock (mean-field) level and find
\begin{equation}
    \delta E^{\xi}=\sum_{\xi_1} \left(\tilde{W}^{\xi \xi_1 \xi_1\xi}_{0}+\tilde{W}^{\xi \xi_1\xi\xi_1}_{0}\right)n_x^{\xi_1}=\frac{d^{\xi}e_0^2}{\epsilon_0 \epsilon^{\perp}_{\mathrm{TMD}}}n_x \ ,
    \label{energyren}
\end{equation}
where the first term reflects direct exciton-exciton interactions and the second term is due to exciton exchange \cite{ciuti1998role}, with the interaction matrix element $\tilde{W}^{\xi_1\xi_2\xi_3\xi_4}_{\bm{q}}\equiv \delta_{\xi_1, \xi_4}\delta_{\xi_2, \xi_3}\tilde{W}^{\xi_1 \xi_2}_{\bm{q}}$ (cf. Eq. \eqref{eqham}). The interaction matrix elements are evaluated in the long wavelength limit, such that the energy renormalization becomes momentum-independent. This is well justified when the exciton distribution is strongly peaked around small centre-of-mass momenta, i.e. at lower temperatures. 
The interaction strength is weighted by the valley-specific hybrid exciton density $n^{\xi}_x=\frac{1}{A}\sum_{\xi, \bm{Q}} N^{(hX) \xi}_{\bm{Q}}$, and the exciton occupation is estimated by a thermalized Boltzmann distribution $N^{(hX) \xi}_{\bm{Q}}\sim n_x \mathrm{exp}(-E^{(hX) \xi}_{\bm{Q}}/(k_B T))$ such that the energy renormalization scales linearly with the total exciton density $n_x=\sum_{\xi} n^{\xi}_x$. This allows us to absorb the exciton-exciton interaction strength and relative occupations in an effective valley-dependent dipole length $d^{\xi}$. In this way, the energy renormalization of a single exciton species $\xi$ is completely characterized by its effective dipole length. A detailed derivation of Eq. \eqref{energyren} and the relevant (electric field-dependent) valley-specific dipole lengths are found in Supplementary Section IV. 
\newline 
\begin{figure}[t!]
    \centering
    \includegraphics[width=\columnwidth]{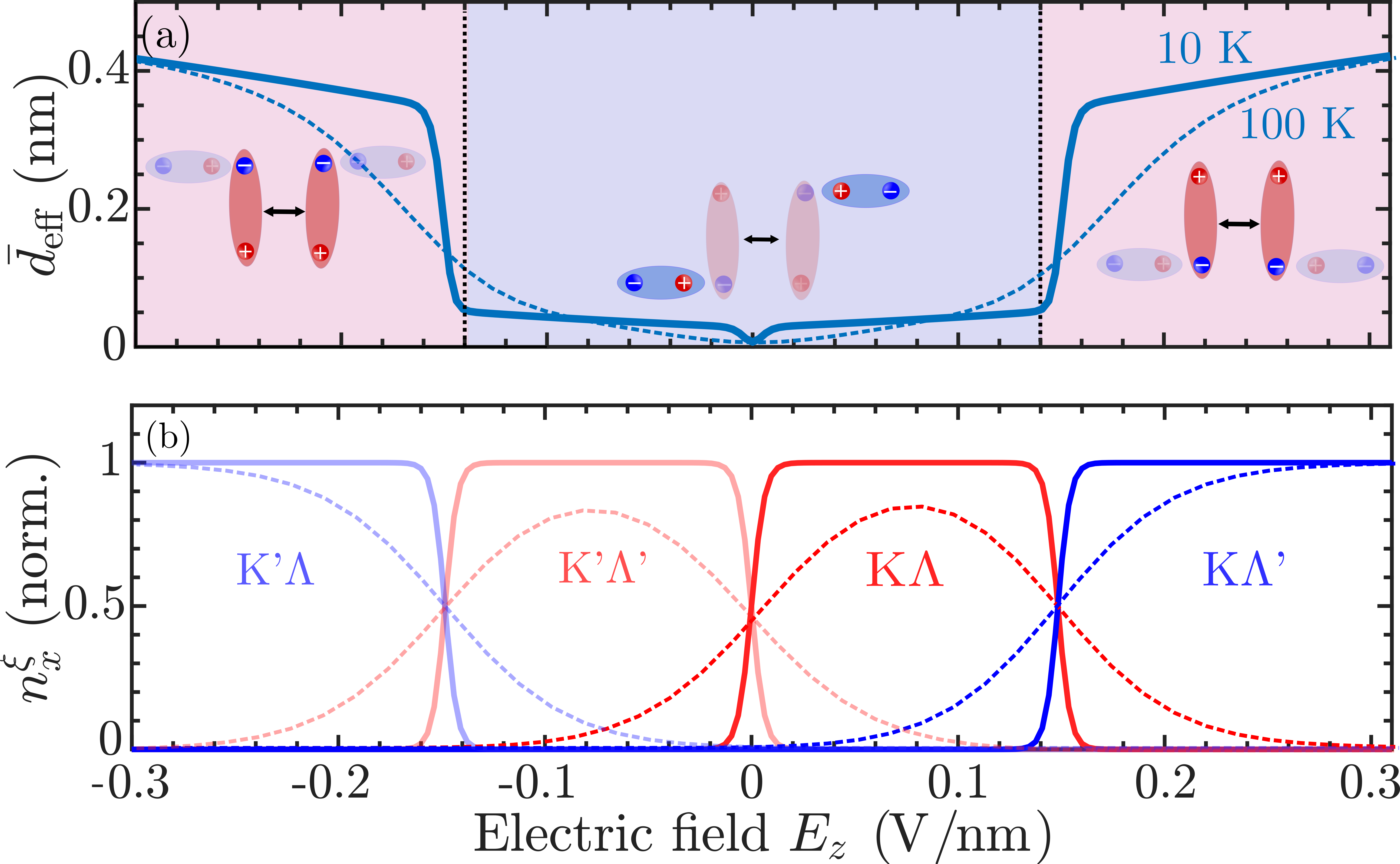}\caption{Dipole length of hybrid excitons. \textbf{(a):} Average dipole length $\bar{d}_{\mathrm{eff}}$ as a function of electric field revealing a drastic increase of the dipole length for $|E_z|>0.15$ V/nm at low temperatures ($T=10$ K). This is explained by a large valley-specific dipole length $d^{\xi}_{\mathrm{eff}}$ of mostly interlayer-like $\xi=$K$\Lambda$'/K'$\Lambda$ excitons and a predominant occupation $n^{\xi}_x$ of these excitons at elevated electric fields shown in \textbf{(b)}). The average dipole length and valley occupations at $T=100$ K are shown as dashed lines.}
    \label{fig3dipoles}
\end{figure}
Furthermore, we define an \emph{average} effective dipole length of the exciton gas, $\bar{d}_{\mathrm{eff}}=n_x^{-1}\sum_{\xi} d^{\xi}n^{\xi}_x$. This quantity is crucial to access macroscopic transport properties, as further discussed in the next section.  
The average effective dipole length is presented in Fig. \ref{fig3dipoles}(a) as function of electric field, together with the normalized valley-specific exciton density in Fig. \ref{fig3dipoles}(b). At low electric fields, the exciton occupation is shared between the energetically degenerate K$\Lambda$ and K'$\Lambda$' states (cf. also Fig. \ref{schematicfig}(d)). These excitons interact weakly via dipolar interactions and combined with their attractive mutual interaction this results in suppressed average effective dipole lengths, $\bar{d}_{\mathrm{eff}}\approx{0.01}$ nm. In contrast, at elevated electric fields $|E_z|>0.15$ V/nm, K$\Lambda$'/K'$\Lambda$ excitons are found to give rise to large effective dipole lengths $\bar{d}_{\mathrm{eff}}\approx{0.4}$ nm, reflecting their large occupation (Fig. \ref{fig3dipoles}(b)) as well as their large interaction strength (Fig. \ref{matrixelement}(b)). Here, we remark that the extracted effective dipole length at large electric fields can be compared with the dipole length of a pure interlayer exciton, $d_{IX}$, here assumed to be equal to the TMD layer thickness, $d_{\mathrm{TMD}}=0.65$ nm (Supplementary Section I). In particular, it holds that $\bar{d}_{\mathrm{eff}}=d_ {IX}|C^{K\Lambda'}_{IX}|^4\approx{0.4}$ nm, i.e. the effective dipole length is obtained by weighting the pure interlayer exciton dipole length by the interlayer component of the mixing coefficient. 
Furthermore, we note that the transition between the low-dipole regime in which K$\Lambda$ and K'$\Lambda$' excitons are prevalent and the high-dipole regime dominated by K$\Lambda$' or K'$\Lambda$ excitons can be tuned by raising the temperature, cf. the dashed curve in Fig. \ref{fig3dipoles}(a) displaying the average effective dipole length at $T=100$ K. This is a consequence of intralayer-like and interlayer-like exciton states being simultaneously populated at high temperatures. Considering the case of $T=100$ K, there exists a sizable occupation of K$\Lambda$/K'$\Lambda$' excitons at finite electric fields and a large electric field ($|E_z|\approx{0.3}$ V/nm) is therefore required for the high-dipole (K$\Lambda$'/K'$\Lambda$) regime to be reached (cf. dashed lines in Fig. \ref{fig3dipoles}(b)).  

\begin{figure}[t!]
    \centering
    \includegraphics[width=0.9\columnwidth]{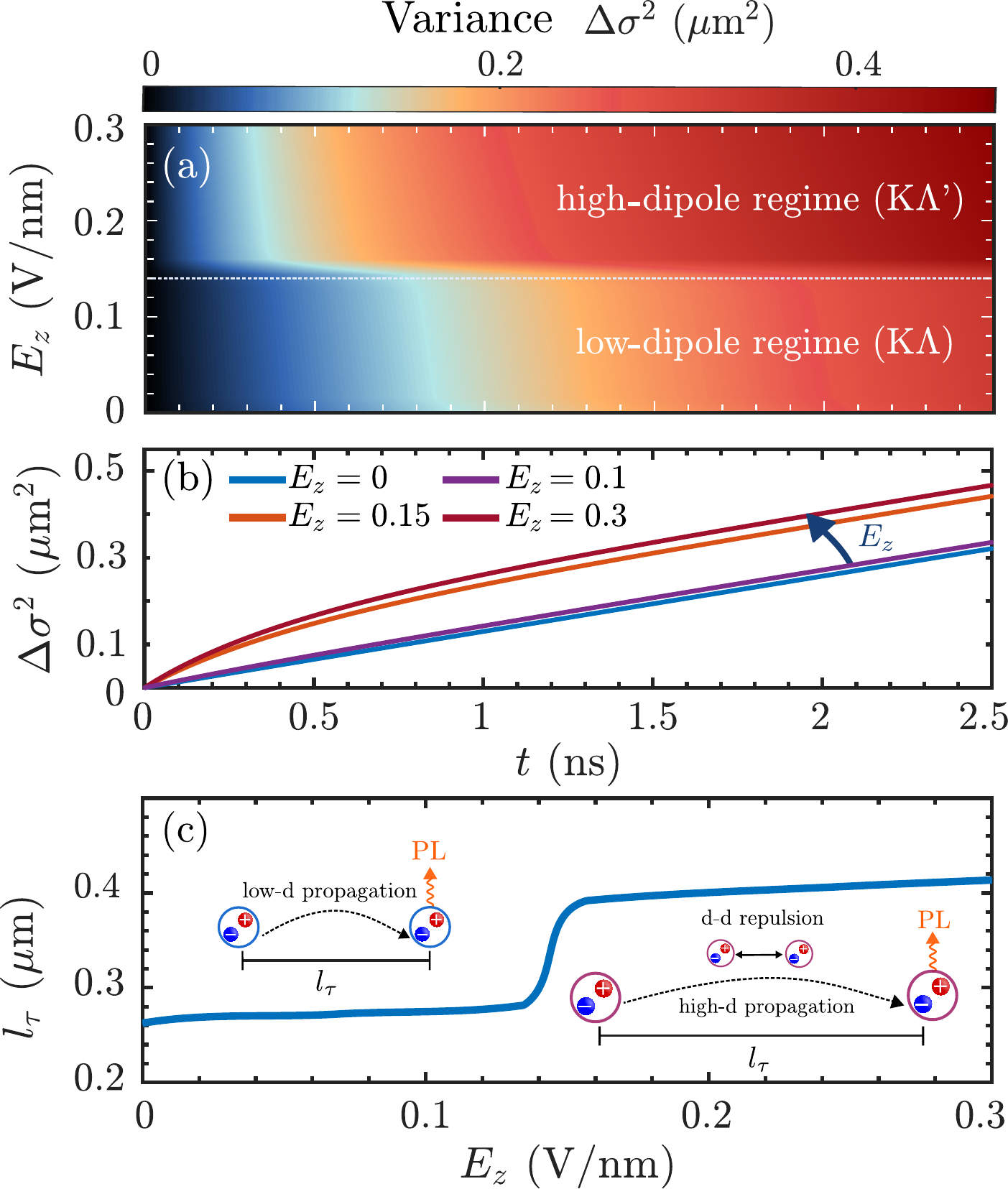}
\caption{Tunability of hybrid exciton diffusion. \textbf{(a):} Electric-field dependent and time-dependent variances $\Delta\sigma^2=\sigma^2_t-\sigma_0^2$ for the exciton density $n(\mathbf{r},t)$, revealing highly anomalous diffusion for electric fields $E_z>0.15$ V/nm at $T=10$ K. The dashed white line indicates the transition from the low-dipole regime (dominated by K$\Lambda$ excitons) to the high-dipole regime (dominated by K$\Lambda$' excitons). \textbf{(b):} Time-dependent variances for different electric fields. For low electric fields ($E_z\leq0.15$ V/nm) conventional diffusion is observed, while at higher electric fields the variance varies non-linearly with time---a hallmark of anomalous diffusion. \textbf{(c): } Diffusion length $l_{\tau}$ as a function of electric field. At large electric fields strong dipole-dipole repulsion between K$\Lambda$' excitons is present, resulting in significantly increased diffusion lengths. }
    \label{diffusionfig}
\end{figure}
\section*{Anomalous hybrid exciton transport}
For a spatially dependent exciton density $n_x\rightarrow n(\bm{r})$, the density-dependent energy renormalization due to repulsive exciton-exciton interactions gives rise to a drift force $-\nabla (\delta E(\bm{r}))$. The latter drags excitons away from the excitation spot \cite{ivanov2002quantum, sun2021excitonic, yuan2020twist} in analogy to exciton funneling in strain potentials \cite{rosati2021dark}. This can be described by the two-dimensional drift-diffusion equation for the exciton density:
\begin{equation}
    \partial_t n(\bm{r}, t)=D\nabla^2 n+\mu\nabla\cdot (\nabla (\delta\bar{E}) n)-\frac{n}{\tau} \ , 
    \label{spatiotemp}
\end{equation}
which is derived using the Wigner function formalism \cite{erkensten2022microscopic, hess1996maxwell}. Here, $D$ is the diffusion coefficient governing the free propagation of excitons, $\mu=\frac{D}{k_B T}$ is the exciton mobility, $T$ being temperature and $\tau$ is the exciton life-time. The second term in Eq. \eqref{spatiotemp} is the drift-term dictated by the averaged energy renormalization $\delta \bar{E}(n(\mathbf{r},t))\equiv \frac{\bar{d}_{\mathrm{eff}}e_0^2}{\epsilon_0 \epsilon^{\perp}_{\mathrm{TMD}}}n(\mathbf{r}, t)$, in which the average effective dipole length $\bar{d}_{\mathrm{eff}}$ crucially enters (Fig. \ref{fig3dipoles}(a)). 
In order to arrive at Eq. \eqref{spatiotemp}, we assumed that all exciton states $\xi$ which contribute to the total exciton population $n=\sum_{\xi} n^{\xi}$ diffuse with the same diffusion coefficient $D^{\xi}\approx{D}$ and that the total population is in thermal equilibrium with the lattice.  
The first assumption is reasonable since the diffusion coefficient is mainly determined by the effective exciton mass, which is the same in the considered states.
The slow thermal equilibration of the exciton gas at low temperatures can, in principle, influence the diffusion dynamics \cite{rosati2020negative}, but these effects have not been observed for the exciton diffusion in van der Waals heterostructures even at cryogenic temperatures \cite{sun2021excitonic}.

We now make use of the strong tunability of the dipole length to show that also the exciton transport can be tuned with respect to electric fields. By numerically solving Eq. \eqref{spatiotemp} we obtain a microscopic access to the spatiotemporal dynamics of excitons. We initialize the exciton density as a typical Gaussian-shaped laser pulse, i.e.  $n(x,y,0)=n_0\mathrm{exp}(-(x^2+y^2)/\sigma_0^2)$ with the initial spot size $\sigma^2_0=1$ $\mu\mathrm{m}^2$ and set the temperature $T=10$ K. The inital exciton density is set to $n_0=10^{12}$ $\mathrm{cm}^{-2}$, such that the drift due to exciton-exciton interactions becomes important and Boltzmann distributions can be used to model the spatiotemporal dynamics of excitons \cite{erkensten2022microscopic}. The considered initial exciton density is below the exciton Mott transition, which is estimated to occur at densities $\sim 7\cdot 10^{12}$ $\mathrm{cm}^{-2}$ in WSe$_2$ homobilayers \cite{siday2022ultrafast} and we neglect the impact of free carriers on the transport \cite{PhysRevB.99.155306}. Moreover, we assume the diffusion coefficient $D=0.3$ $\mathrm{cm}^2/\mathrm{s}$ and the exciton life-time $\tau=500$ ps as obtained from a recent experiment on the same homobilayer \cite{tagarelli23}. Note that there is no moir{\'e} potential which could trap excitons and slow down their propagation \cite{choi2020moire, yuan2020twist, PhysRevLett.126.106804, knorr2022exciton}, as we are considering untwisted homobilayers with no lattice mismatch. 

In Fig. \ref{diffusionfig}(a), the time- and electric-field dependent variance $\Delta\sigma^2=\sigma_t^2-\sigma_0^2$ is shown for the case of naturally stacked hBN-encapsulated WSe$_2$ homobilayers, revealing a significant broadening of the exciton spatial distribution at electric fields $E_z>0.15$ V/nm, corresponding to the high-dipole regime (cf. Fig. \ref{fig3dipoles}(a)). The transition from the low-dipole to the high-dipole regime results in highly non-linear exciton transport, cf. Fig. \ref{diffusionfig}(b). In the low-dipole regime (Fig. \ref{fig3dipoles}(a)), excitons are not affected by any drift and the width of the distribution varies approximately linearly with time, i.e. $\Delta\sigma^2=4D t$ according to Fick's law. In contrast, in the high-dipole regime, the exciton drift is highly efficient leading to a super-linear dependence on the variance with respect to time---a hallmark of anomalous diffusion \cite{yuan2020twist}. 
Finally, given the fully time-resolved broadening of the exciton distribution, we extract a time-independent measure of the exciton transport, i.e. the experimentally tractable diffusion length $l_{\tau}=2\sqrt{D_{\tau} \tau}=\sqrt{\sigma^2_{\tau}-\sigma_0^2}$ \cite{PhysRevLett.120.207401}. This quantity is a measure for how far away from the excitation spot the excitons propagate before recombining and should thus be enhanced with the exciton drift due to dipole-dipole repulsion (Fig. \ref{diffusionfig}(c)). We obtain diffusion lengths in the submicrometer range, concretely $0.25$ $\mu\mathrm{m}$ and $0.40 \ \mu\mathrm{m}$ in the low- and high-dipole regime, respectively, which are similar to diffusion lengths obtained from previous transport measurements on TMD monolayers and bilayers \cite{cadiz2018exciton, choi2020moire}.
We note that excitons in MoSe$_2$/hBN/WSe$_2$ heterostructures have been reported to exhibit longer diffusion lengths of $1-2$ $\mu\mathrm{m}$ \cite{sun2021excitonic}, since excitons in these structures are of purely interlayer character and exhibit enhanced dipole moments due to the hBN spacer.
Overall, we reveal a remarkable tunability of the diffusion length in the considered WSe$_2$ bilayers with electric fields and we find that the transport of hybrid excitons can be electrically controlled, which is of importance for the realization of exciton-based optoelectronic devices \cite{unuchek2018room, ciarrocchi2019polarization}. 

\section*{Conclusions}
Our work sheds light on the impact of electric fields on hybrid exciton-exciton interactions in transition-metal dichalcogenide bilayers. We highlight the presence of hybrid excitons in naturally-stacked WSe$_2$ homobilayers and find that the energetically lowest exciton state in these structures can be tuned by applying an out-of-plane electric field. The nature of the lowest state varies from mostly intralayer to mostly interlayer character, resulting in a low-dipole and a high-dipole regime at small and large electric fields, respectively. The latter is characterized by strong interactions between hybrid excitons due to the efficient dipole-dipole repulsion. The electrical tunability of the interactions has also direct consequences on exciton transport. In particular, we predict that the transition from low- to high-dipole regime is accompanied by the emergence of anomalous exciton diffusion, which is a characteristic fingerprint of strong dipole-dipole repulsion. The insights obtained from our material-specific and predictive many-particle theory can be used to guide experiments to measure the tunability of hybrid exciton-exciton interactions in atomically thin semiconductors. Furthermore, our study provides tools for investigating the impact of electrically tunable exciton-exciton interactions on other exotic phenomena in semiconductor moir{\'e} materials such as exciton condensation and superfluidity. \\
\section*{Acknowledgments}
We thank Roberto Rosati (Philipps-Universität Marburg) for useful discussions. This project has received funding from  Deutsche Forschungsgemeinschaft via CRC 1083 and the project 512604469 as well as from the European Unions Horizon 2020 research and innovation programme under grant agreement no. 881603 (Graphene Flagship). The Kis group received funding from the Swiss National Science Foundation (grants no. 164015, 177007, 175822, 205114), and the Marie Curie Sklodowska ITN network “2-Exciting” (grant no. 956813).

\end{document}


\renewcommand{\theequation}{S\arabic{equation}}
\renewcommand{\thesection}{\Roman{section}}
\preprint{APS/123-QED}

\title{Supplementary Material for \--- Electrical tunability of hybrid exciton-exciton interactions in transition-metal dichalcogenide bilayers}
\title{\textbf{Supplementary Material for} \\ Electrically tunable dipolar interactions between layer-hybridized excitons}
\author{Daniel Erkensten$^1$, Samuel Brem$^2$, Ra\"ul Perea-Caus\'in$^1$, Joakim Hagel$^1$, Fedele Tagarelli$^3$, Edoardo Lopriore$^3$, Andras Kis$^3$and Ermin Malic$^{2,1}$}
\affiliation{$^1$Department of Physics, Chalmers University of Technology, 41296 Gothenburg, Sweden}
\affiliation{$^2$Department of Physics, Philipps-Universit{\"a}t Marburg, 35037 Marburg, Germany}
\affiliation{$^3$Institute of Electrical and Microengineering, École Polytechnique Fédérale de Lausanne (EPFL), Lausanne, Switzerland}
\maketitle 
\date{\today}
\section{Hybrid exciton landscape in TMD bilayers}
\label{energysec}
In here, we discuss how the exciton landscape in TMD bilayers is obtained within our theoretical framework, taking into account the effect of layer-hybridization. The starting-point is a two-dimensional system containing pure intra- and interlayer excitons. The intra- and interlayer exciton binding energies and wave functions in TMD bilayers are obtained from solving the bilayer Wannier equation \cite{kira2006many}
\begin{equation}
   \frac{\hbar^2 \bm{k}^2}{2m^{\xi L}_{\mathrm{red}}}\varphi^{\xi L}_{n, \bm{k}}-\sum_{\bm{q}}V^{c_{l_e}v_{l_h}}_{\bm{q}}\varphi^{\xi L}_{n, \bm{k}+\bm{q}}=E^{\xi L}_{n, \mathrm{bind}} \varphi^{\xi L}_{n, \bm{k}} \ , 
\end{equation}
where $\varphi^{\xi L}_{n,\bm{k}}$ is the excitonic wave function in state $n=1s,2s...$, valley $\xi=(\xi_e, \xi_h)$, and layer $L=(l_e, l_h)$ and $E^{\xi L}_{n, \mathrm{bind}}$ is the exciton binding energy. Here, the reduced exciton mass $m^{\xi L}_{\mathrm{red}}=\frac{m^{\xi_e l_e} m^{\xi_h l_h} }{m^{\xi_h l_h}+m^{\xi_e l_e}}$, as well as the screened electron-hole Coulomb interaction $V^{c_{l_e}v_{l_h}}_{\bm{q}}$ enter. The valley-specific electron (hole) masses $m^{\xi_e l_e}$ ($m^{\xi_h l_h}$) are obtained from DFT calculations \cite{kormanyos2015k}. When evaluating the Coulomb matrix elements we explicitly include the finite thickness of the TMD layers as well as the as the dielectric environment via a generalized Keldysh screening \cite{ovesen2019interlayer}. In this work, we explicitly include hybridization of intra ($X$)- and interlayer ($IX$) excitons. In particular, the four possible intra- and interlayer exciton states (here expressed as $L\equiv IX_1, IX_2, X_1, X_2$ focusing on the energetically lowest $n=1s$ transitions such that the exciton index can be omitted) are generally coupled by electron/hole tunneling. The resulting hybrid exciton states are obtained from diagonalizing the following exciton Hamiltonian \cite{hagel2021exciton, brem2020hybridized}
\begin{equation}
    H_{x,0}=\sum_{\xi, L, \bm{Q}}E^{\xi}_{L, \bm{Q}}X^{\dagger \xi}_{L, \bm{Q}}X^{\xi}_{L, \bm{Q}}+\sum_{\xi, L, L', \bm{Q}}T^{\xi}_{LL'}X^{\dagger \xi}_{L, \bm{Q}}X^{\xi}_{L', \bm{Q}} \ , 
    \label{tunneling}
\end{equation}
containing the exciton centre-of-mass dispersion $E^{\xi}_{L,\bm{Q}}=\frac{\hbar^2 \bm{Q}^2}{2M^{\xi L}}+E^{\xi L}_{\mathrm{bind}}+\Delta^{\xi L}$, $M^{\xi L}=m^{\xi_h l_h}+m^{\xi_e l_e}$ being the total exciton mass, $X^{(\dagger)}$ being excitonic and bosonic ladder operators and $\Delta^{\xi L}$ is the valley-specific band gap. The free Hamiltonian also contains a tunneling contribution which takes into account the tunneling of electrons and holes between different layers ($l_e\neq l'_e$ or $l_h\neq l'_h$) via the excitonic tunneling matrix element
\begin{equation}
    T^{\xi}_{LL'}=F^{\xi}_{LL'}[T^{c\xi_e}_{l_el'_e}\delta_{l_h, l'_h}(1-\delta_{l'_e, l_e})-T^{v\xi_h}_{l_hl'_h}\delta_{l_e, l'_e}(1-\delta_{l'_h, l_h})] \ .
\end{equation}
The excitonic tunneling matrix element crucially depends on electron and hole tunneling strengths, $T^{c\xi_e}_{l_e l'_e}$ and $T^{v\xi_h}_{l_h l'_h}$ respectively, as well as exciton wave function overlaps $F^{\xi }_{LL'}=\sum_{\bm{k}}\varphi^{*\xi L}_{\bm{k}}\varphi^{\xi L'}_{\bm{k}}$. The electron and hole tunneling strengths are obtained from \emph{ab-initio} calculations and are reported in Ref. \cite{hagel2021exciton} for common TMD bilayers. For the considered case of 2H-stacked WSe$_2$ homobilayers we adopt the tunneling strengths $T^{c K}=0$, $T^{v K}=66.9$ meV and $T^{c \Lambda}=236.6$ meV for the most relevant K/K' and $\Lambda$/$\Lambda$' valleys in this structure. Note that the electronic tunneling matrix elements are generally stacking- and momentum-dependent, however in this work we focus on naturally stacked ($H^{h}_h$) homobilayers, and evaluate the matrix elements at the high-symmetry points. The Hamiltonian in Eq. \eqref{tunneling} is now diagonalized via the basis transformation \cite{brem2020hybridized}
\begin{equation}
    X^{\dagger \xi}_{L, \bm{Q}}=\sum_{\eta} C^{\xi\eta }_{L}(\bm{Q})Y^{\dagger \xi}_{\eta, \bm{Q}} \ , 
    \label{trans}
\end{equation}
where $Y^{(\dagger)}$ is a new set of \emph{hybrid} exciton operators and $C^{\xi\eta }_{L}(\bm{Q})$ is the \emph{mixing coefficient} determining the relative intra/interlayer content of the hybrid exciton, enabling us to define a hybrid exciton state as $|hX_{\eta} \rangle=\sum_{i=1,2} (C^{\eta}_{X_i}|X_i\rangle +C^{\eta}_{IX_i}|IX_i\rangle)$ with $\sum_{i=1,2}(|C^{\eta}_{X_i}|^2+|C^{\eta}_{IX_i}|^2)=1$ for a fixed hybrid exciton state $\eta$. The mixing coefficients are obtained from solving the following hybrid eigenvalue problem\cite{brem2020tunable}
\begin{equation}
    E^{\xi}_{L, \bm{Q}}C^{\xi \eta}_{L}(\bm{Q}) +\sum_{L'}T^{\xi}_{LL'}C^{\xi \eta}_{L'}(\bm{Q}) = E^{(hX) \xi}_{\eta, \bm{Q}}C^{\xi \eta}_{L}(\bm{Q}) \ , 
    \label{eq2}
\end{equation}
introducing the hybrid exciton eigenenergy $E^{(hX) \xi}_{\eta, \bm{Q}}$. We can now express the exciton Hamiltonian above in the hybrid basis such that $H_{x,0}\rightarrow \tilde{H}_{x,0}$ with
\begin{equation}
    \tilde{H}_{x,0}=\sum_{\xi, \eta, Q}E^{(hX)\xi}_{\eta, \bm{Q}}Y^{\dagger \xi}_{\eta, \bm{Q}}Y^{\xi}_{\eta, \bm{Q}} \ . 
\end{equation}
By solving the eigenvalue problem in Eq. \eqref{eq2} we get microscopic access to the full hybrid exciton landscape in TMD bilayers. Furthermore, we investigate the impact of an electric field on the hybrid exciton landscape. This is done by exploiting the electrostatic Stark shift of interlayer resonances, i.e. by taking $E^{\xi}_{L=IX, \bm{Q}}\rightarrow E^{\xi}_{L=IX, \bm{Q}}+ \Delta E$, with $\Delta E=\pm d e_0 E_z$, where $d\approx{0.65}$ nm is the dipole length (here assumed to be the same as the TMD layer thickness \cite{laturia2018dielectric}),$e_0$ is the electric charge and $E_z$ is the out-of-plane electric field \cite{hagel2022electrical}. In this way, hybrid exciton eigenenergies and mixing coefficients become tunable with respect to electric fields. In Table \ref{energytable}, we report the exciton hybrid energies and intralayer and interlayer mixing coefficients for hybrid excitons composed by electrons in the $\xi_e=$ K, K', $\Lambda$, $\Lambda$' valleys and holes in the $\xi_h=$ K, K' valleys in 2H-stacked hBN-encapsulated WSe$_2$ homobilayers. The energies are given relative to the intralayer A exciton energy and the electric fields $E_z=0, \pm 0.3$ V/nm are considered. 
\begin{table}[h!]
\begin{tabular}{|c|c|c|c|c|c|c|c|c|c|}
Exciton  & \multicolumn{3}{c|}{Energy $E-E_{X_A}$(meV)}& \multicolumn{3}{c|}{Intralayer component $|C_X|^2$} & \multicolumn{3}{c|}{Interlayer component $|C_{IX}|^2$} \\ \hline
$\xi=(\xi^h, \xi^e$)  & $E_z=-0.3$ & $E_z=0$ & $E_z=0.3$ &  $E_z=-0.3$ &  $E_z=0$ & $E_z=0.3$ & $E_z=-0.3$ & $E_z=0$ & $E_z=0.3$  \\ \hline 
K$\Lambda$ & -123 & \textbf{-159} & -209 & 0.85 & \textbf{0.77} & 0.61 & 0.15 & \textbf{0.23} & 0.39 \\ 
 K’$\mathrm{\Lambda}$' & -209 & \textbf{-159} & -123 & 0.61 & \textbf{0.77} & 0.85 & 0.39 & \textbf{0.23} & 0.15 \\
K$\mathrm{\Lambda}$' & -12 & -117 & \textbf{-258} & 0.58 & 0.36 & \textbf{0.2} & 0.42 & 0.64 & \textbf{0.8} \\ 
K’$\mathrm{\Lambda}$ & \textbf{-258} & -117 & -12 & \textbf{0.2} & 0.36 & 0.58 & \textbf{0.8} & 0.64 & 0.42 \\
KK & -5 & 0 & -179 & 0.96 & 1 & 0.01 & 0.04 & 0 & 0.99 \\ 
K’K’ & -179 & 0 & -5 & 0.01 & 1 & 0.96 & 0.99 & 0 & 0.04 \\ 
KK’ & -58 & -53 & -137 & 0.96 & 1 & 0.01 & 0.04 & 0 & 0.99 \\ 
K’K & -137 & -53 & -58 & 0.01 & 1 & 0.96 & 0.99 & 0 & 0.04 \\ 
\end{tabular}
\caption{Exciton landscape in hBN-encapsulated 2H-stacked WSe$_2$ homobilayers. We provide the valley-specific energies, intralayer components and interlayer components of hybrid excitons for three different values on the electric field, $E_z=0, \pm 0.3$ V/nm.  The energetically lowest transitions for each electric field are marked in bold and energies are given relative to the KK intralayer A exciton energy ($E_{X_A}$). For vanishing electric fields we find that the K$\mathrm{\Lambda}$ and K'$\Lambda$' states represent the energetically lowest states. }
\label{energytable}
\end{table}\\
Note that the K$\Lambda$ and K'$\Lambda$' exciton states are energetically degenerate at vanishing electric fields. This is a consequence of the H-type stacking, where the individual TMD layers are rotated 180 degrees with respect to each other such that the spin-orbit coupling in one of the layers is inverted. Moreover, these states can be expressed as $|K\Lambda \rangle= C^{K\Lambda}_{X_1}|X_1\rangle +C^{K\Lambda}_{IX_1}|IX_1\rangle$ and $|K'\Lambda' \rangle= C^{K'\Lambda'}_{X_2}|X_2\rangle +C^{K'\Lambda'}_{IX_2}|IX_2\rangle$ such that each of the states only mixes contributions from a single intralayer and a single interlayer exciton species. Hence, it follows that the K$\Lambda$ and K'$\Lambda$' hX carry opposite out-of-plane dipole moments via their interlayer components, and therefore the energy of these states shifts in opposite directions under the application of an electric field (cf. Table \ref{energytable}).

\section{Hybrid exciton-exciton interaction Hamiltonian} 
\label{interactionsec}
 Here, we provide a microscopic derivation of the hybrid exciton-exciton interaction Hamiltonian. The starting-point is the bilayer carrier-carrier Hamiltonian: 
 \begin{equation}
     H_{c-c}=\frac{1}{2}\sum_{\lambda^{(')}, \xi^{(')}, l^{(')}}V^{\lambda_l \lambda'_{l'}}_{\bm{q}}\lambda^{\dagger}_{\xi, l, \bm{k+q}}\lambda'^{\dagger}_{\xi', l', \bm{k}'-\bm{q}}\lambda'_{\xi', l', \bm{k}'}\lambda_{\xi, l, \bm{k}} \ , 
     \label{hamelectronic}
 \end{equation}
where $\lambda^{(')}=(c,v)$, $\xi$, and $l^{(')}$ are the band, valley, and layer indices, respectively. Here, the operators $\lambda^{(\dagger)}$ annihilate (create) carriers in band $\lambda$. Moreover, we note that $V^{\lambda_l\lambda_{l'}}_{\bm{q}}$ describes an intraband intralayer Coulomb interaction if $l=l'$ and an interlayer Coulomb interaction if $l\neq l'$. Furthermore, we consider the long-range part of the Coulomb interaction such that $V^{\lambda_l\lambda'_l}_{\bm{q}}\approx{\frac{e^2_0}{2\epsilon_0 A|\bm{q}|\epsilon_{\mathrm{intra}, \bm{q}}}}$ and $V^{\lambda_l\lambda'_{\bar{l}}}_{\bm{q}}\approx{\frac{e^2_0}{2\epsilon_0 A |\bm{q}| \epsilon_{\mathrm{inter}, \bm{q}}}}$ ($l\neq \bar{l}$). The intra- and interlayer dielectric functions $\epsilon_{\mathrm{intra}, \bm{q}}$ and $\epsilon_{\mathrm{inter}, \bm{q}}$ can be found in the Supplementary Material of Ref. \cite{erkensten2022microscopic}. Interband Coulomb interactions, which give rise to electron-hole exchange \cite{katsch2019theory} or Auger recombination \cite{erkensten2021exciton}, are not expected to contribute significantly to experimentally accessible density-dependent energy renormalizations (Supplementary Section IV) and are therefore neglected in this work.

Given the carrier-carrier Hamiltonian, we now proceed as follows: \textbf{i)} we find the equation of motion for the intervalley polarisation $\langle P^{\dagger \xi_e l_e, \xi_h l_h}_{\bm{k}_1+\bm{Q}, \bm{k}_1}\rangle \equiv \langle c^{\dagger}_{\xi_e, l_e, \bm{k}_1+\bm{Q}}v_{\xi_h, l_h, \bm{k}_1}\rangle$, \textbf{ii)} transform the equation of motion to the exciton basis \cite{katsch2018theory}, \textbf{iii)} make an ansatz for the exciton-exciton interaction Hamiltonian and compute the equation of motion for the polarisation in the exciton picture, \textbf{iv)} read off the exciton-exciton interaction matrix element such that the results from steps \textbf{ii)} and \textbf{iii)} coincide. Starting with the first step \textbf{i)}, we obtain the equation of motion for the polarisation directly from the Heisenberg equation of motion \cite{kira2006many}. Including only the Coulomb contributions from Eq. \eqref{hamelectronic} we obtain 
 \begin{align}
     i\hbar \frac{d}{dt}\langle P^{\dagger{\xi_el_e, \xi_hl_h}}_{\bm{k}_1+\bm{Q}, \bm{k}_1}\rangle
     &=\frac{1}{2}\sum_{\bm{k}, \bm{q}, l, \xi} \bigg( V^{v_{l_h} v_l}_{\bm{q}}(\langle c^{\dagger}_{\xi_e,l_e,  \bm{k}_1+\bm{Q}}v_{\xi, l, \bm{k}}v_{\xi_h, l_h, \bm{k}_1-\bm{q}}v^{\dagger}_{\xi, l, \bm{k}-\bm{q}}\rangle-\langle c^{\dagger}_{\xi_e,l_e,  \bm{k}_1+\bm{Q}}v_{\xi_h, l_h, \bm{k}_1-\bm{q}}v_{\xi, l, \bm{k}}v^{\dagger}_{\xi, l, \bm{k}-\bm{q}}\rangle)\nonumber\\\nonumber
     & \ \ \ \ \ +V^{c_l c_{l_e}}_{\bm{q}}(\langle  c^{\dagger}_{\xi, l,\bm{k+q}} v_{\xi_h, l_h, \bm{k}_1}c^{\dagger}_{\xi_e, l_e, \bm{k}_1+\bm{Q}-\bm{q}}c_{\xi, l, \bm{k}}\rangle-\langle c^{\dagger}_{\xi_e, l_e, \bm{k}_1+\bm{Q}-\bm{q}}v_{\xi_h, l_h, \bm{k}_1}c^{\dagger}_{\xi, l,\bm{k+q}}c_{\xi, l, \bm{k}}\rangle)\\\nonumber
     & \ \ \ \ \ + V^{v_l c_{l_e}}_{\bm{q}}(\langle c^{\dagger}_{\xi_e, l_e, \bm{k}_1+\bm{Q}-\bm{q}}v_{\xi_h, l_h, \bm{k}_1}v_{\xi, l, \bm{k}}v^{\dagger}_{\xi, l,\bm{k+q}}\rangle-\langle c^{\dagger}_{\xi_e, l_e, \bm{k}_1+\bm{Q}-\bm{q}}v_{\xi, l, \bm{k}}v_{\xi_h, l_h, \bm{k}_1}v^{\dagger}_{\xi, l,\bm{k+q}}\rangle)\\
          & \ \ \ \ \ +  V^{v_{l_h}c_l}_{\bm{q}}(\langle c^{\dagger}_{\xi_e,l_e,  \bm{k}_1+\bm{Q}}v_{\xi_h, l_h, \bm{k}_1-\bm{q}}c^{\dagger}_{\xi, l, \bm{k}-\bm{q}}c_{\xi, l, \bm{k}}\rangle-\langle c^{\dagger}_{\xi, l, \bm{k}-\bm{q}}v_{\xi_h, l_h, \bm{k}_1-\bm{q}}c^{\dagger}_{\xi_e,l_e,  \bm{k}_1+\bm{Q}}c_{\xi, l, \bm{k}}\rangle)\bigg) \ .
          \label{eqelec}
 \end{align}
 Next, we transform the entire equation above to the excitonic basis and make use of the pair operator expansions \cite{katsch2018theory}
 \begin{equation}
     c^{\dagger}_{\xi_e, l_e, \bm{k}}c_{\xi'_e, l'_e, \bm{k'}}\approx{ \sum_{\xi''_h, l''_h, \bm{k}''} P^{\dagger \xi_e l_e, \xi''_h l''_h}_{\bm{k}, \bm{k}''}P^{\xi'_e l'_e, \xi''_h l''_h}_{\bm{k}', \bm{k}''} } \ ,  v_{\xi_h, l_h, \bm{k}}v^{\dagger}_{\xi'_h, l'_h, \bm{k'}}\approx{ \sum_{\xi''_e, l''_e, \bm{k}''} P^{\dagger \xi''_e l''_e, \xi_h l_h}_{\bm{k}'', \bm{k}}P^{\xi''_e l''_e, \xi'_h l'_h}_{\bm{k}'', \bm{k}'} } \ ,  
 \end{equation}
 where the pair operators can be further expressed in the exciton basis as $P^{\xi_e l_e, \xi_h l_h}_{\bm{k}, \bm{k}'}=\sum_{n} \varphi^{\xi L}_{n, \beta^{\xi L}\bm{k}+\alpha^{\xi L}\bm{k}'}X^{\xi}_{n, L, \bm{k}-\bm{k}'}$, with $\varphi^{\xi L}_{n, \bm{k}}$ being the exciton wave function (cf. Supplementary Section I) and the compound indices $\xi=(\xi_e, \xi_h)$, $L=(l_e, l_h)$ (such that $l_e=l_h$ corresponds to the intralayer wave function and $l_e\neq l_h$ corresponds to the interlayer wave function). In the following, we will only consider the lowest-lying $n=1s$ exciton states, so that the index $n$ can be omitted. By doing this, the equation of motion Eq. \eqref{eqelec} separates into two parts, a direct part and an exchange part. The second, fourth, fifth and seventh term in Eq. \eqref{eqelec} gives rise to the direct terms reading 
 \begin{align}
     i\hbar \frac{d}{dt}\langle X^{\dagger \xi'}_{L', \bm{Q}}\rangle|_{\mathrm{dir.}} &=\frac{1}{2}\sum_{\bm{q}, \bm{Q}_1, \xi, L}\bigg( V^{c_{l'_e}v_{l_h}}_{\bm{q}}F(\alpha^{\xi'L'}\bm{q})F(\beta^{\xi L}\bm{q})+V^{c_{l_e}v_{l'_h}}_{\bm{q}}F(-\alpha^{\xi'L'}\bm{q})F(-\beta^{\xi L}\bm{q})\label{eqdir}\\\nonumber 
     & - V^{c_{l_e}c_{l'_e}}_{\bm{q}}F(\beta^{\xi'L'}\bm{q})F(-\beta^{\xi L}\bm{q}))-V^{v_{l'_h}v_{l_h}}_{\bm{q}}F(-\alpha^{\xi'L'}\bm{q})F(\alpha^{\xi L}\bm{q})\bigg)\langle X^{\dagger \xi'}_{L', \bm{Q}+\bm{q}}X^{\dagger \xi}_{L, \bm{Q}_1-\bm{q}}X^{\xi}_{L, \bm{Q}_1}\rangle \ , 
 \end{align}
 where we introduced the compound indices $\xi^{(')}=(\xi^{(')}_e, \xi^{(')}_h)$ and $L^{(')}=(l^{(')}_e, l^{(')}_h)$. Here, we also defined the excitonic form factors $F(x^{\xi L}\bm{q})\equiv \sum_{\bm{k}}\varphi^{*\xi L}_{\bm{k}+x^{\xi L}\bm{q}}\varphi^{\xi L}_{\bm{k}}$. We may now construct the corresponding direct exciton-exciton interaction Hamiltonian with 
 \begin{equation}
     H_{x-x}|_{\mathrm{dir.}}=\frac{1}{2}\sum_{\substack{\bm{Q}_1, \bm{Q}_2, \bm{q}\\ \xi, \xi', L, L'}}D^{\xi\xi'}_{L, L', \bm{q}}X^{\dagger \xi'}_{L', \bm{Q}_1+\bm{q}}X^{\dagger \xi}_{L, \bm{Q}_2-\bm{q}}X^{\xi}_{L, \bm{Q}_2}X^{ \xi'}_{L', \bm{Q}_1} \ , 
     \label{hamdir}
 \end{equation}
 with the direct part of the exciton-exciton interaction reading 
 \begin{align}
     D^{\xi\xi'}_{L, L', \bm{q}}&=\frac{1}{2}\bigg(
     V^{c_{l_e}c_{l'_e}}_{\bm{q}}F(\beta^{\xi'L'}\bm{q})F(-\beta^{\xi L}\bm{q}))+V^{v_{l'_h}v_{l_h}}_{\bm{q}}F(-\alpha^{\xi'L'}\bm{q})F(\alpha^{\xi L}\bm{q})\label{dirint}\\
     &-V^{c_{l'_e}v_{l_h}}_{\bm{q}}F(\alpha^{\xi'L'}\bm{q})F(\beta^{\xi L}\bm{q})-V^{c_{l_e}v_{l'_h}}_{\bm{q}}F(-\alpha^{\xi'L'}\bm{q})F(-\beta^{\xi L}\bm{q})\bigg)   \nonumber\ ,
 \end{align}
 such that a commutation of the excitonic Hamiltonian \eqref{hamdir} with the polarisation gives rise to Eq. \eqref{eqdir}. We note that, in the long wavelength limit
 \begin{equation}
     D^{\xi\xi'}_{X_i, X_i, \bm{0}}=D^{\xi\xi'}_{X_i, X_j, \bm{0}}=0 \ ,  D^{\xi\xi'}_{IX_i, IX_i, \bm{0}}=-D^{\xi\xi'}_{IX_i, IX_j, \bm{0}}=\frac{e_0^2}{4A \epsilon_0 }(\frac{d_{1, \mathrm{TMD}}}{\epsilon^{\perp}_{1, \mathrm{TMD}}}+\frac{d_{2, \mathrm{TMD}}}{\epsilon^{\perp}_{2, \mathrm{TMD}}}) \ , i\neq j
     \label{dirspecial}
 \end{equation}
 i.e. we find a vanishing direct interaction between intralayer excitons ($L, L'=X_i$, $i=1,2$) and recover the widely used plate capacitor formula when considering interactions between interlayer excitons ($L, L'=IX_i$, $i=1,2$) as has been previously confirmed in literature \cite{ciuti1998role, erkensten2021exciton, PhysRevB.96.115409}. Here, the material-specific constants $d_{i, \mathrm{TMD}}$ and $\epsilon^{\perp}_{i, \mathrm{TMD}}$ denote individual TMD monolayer thicknesses and out-of-plane components of the TMD dielectric tensors, respectively. In the main manuscript we set $d_{1,\mathrm{TMD}}=d_{2,\mathrm{TMD}}\equiv d_{\mathrm{TMD}}$ and $\epsilon^{\perp}_{1, \mathrm{TMD}}=\epsilon^{\perp}_{2, \mathrm{TMD}}\equiv \epsilon^{\perp}_{\mathrm{TMD}}$ as we are considering a homobilayer. Note that interactions between different interlayer exciton species $IX_i$ and $IX_j$, $i\neq j$ are attractive due to the opposite dipole orientations of these excitons. Now, we consider the remaining terms (i.e. the first, third, fifth and eight terms) in Eq. \eqref{eqelec} and find that these give rise to the following exchange terms
 \begin{align}
     i\hbar \frac{d}{dt}\langle X^{\dagger\xi'}_{L', \bm{Q}}\rangle |_{\mathrm{exch.}}&=\frac{1}{2}\sum_{\substack{\bm{q}, \bm{Q}_1, \bm{k}, \bm{k}'\\ \xi, \tilde{\xi}, \bar{\xi}\\ L, \tilde{L}, \bar{L}} }\bigg( (V^{c_{l_e}c_{l'_e}}_{\bm{k}-\bm{k}'}\varphi^{\xi'L'}_{\bm{k}-\alpha^{\xi'L'}\bm{Q}-\bm{q}}-V^{c_{l_e}v_{l'_h}}_{\bm{k}-\bm{k}'}\varphi^{\xi'L'}_{\bm{k}'-\alpha^{\xi'L'}\bm{Q}-\bm{q}})\times \\\nonumber 
     &\delta^{\xi_h, \xi'_h}_{l_h, l'_h}\delta^{\tilde{\xi}_e, \xi'_e}_{\tilde{l}_e, l'_e}\delta^{\bar{\xi}_e, \xi_e}_{\bar{l}_e, l_e}\delta^{\bar{\xi}_h, \tilde{\xi}_h}_{\bar{l}_h, \tilde{l}_h}\varphi^{*\xi L}_{\bm{k}-\alpha^{\xi L}(\bm{Q}+\bm{q})}\varphi^{*\tilde{\xi}\tilde{L}}_{\bm{k}'-\beta^{\tilde{\xi}\tilde{L}}\bm{q}-\alpha^{\tilde{\xi}\tilde{L}}\bm{Q}_1}\varphi^{\bar{\xi}\bar{L}}_{\bm{k}'-\alpha^{\bar{\xi}\bar{L}}\bm{Q}_1} \\\nonumber 
     &+ (V^{v_{l'_h}v_{l_h}}_{\bm{k}-\bm{k}'}\varphi^{\xi'L'}_{\bm{k}+\beta^{\xi'L'}\bm{Q}+\bm{q}}-V^{c_{l'_e}v_{l_h}}_{\bm{k}-\bm{k}'}\varphi^{\xi'L'}_{\bm{k}'+\beta^{\xi'L'}\bm{Q}+\bm{q}})\times \\\nonumber 
     &\delta^{\xi_e, \xi'_e}_{l_e, l'_e}\delta^{\tilde{\xi}_h, \xi'_h}_{\tilde{l}_h, l'_h}\delta^{\bar{\xi}_e, \tilde{\xi}_e}_{\bar{l}_e, \tilde{l}_e}\delta^{\bar{\xi}_h, \xi_h}_{\bar{l}_h, l_h}\varphi^{*\xi L}_{\bm{k}+\beta^{\xi L}(\bm{Q}+\bm{q})}\varphi^{*\tilde{\xi}\tilde{L}}_{\bm{k}'+\alpha^{\tilde{\xi}\tilde{L}}\bm{q}+\beta^{\tilde{\xi}\tilde{L}}\bm{Q}_1}\varphi^{\bar{\xi}\bar{L}}_{\bm{k}'+\beta^{\bar{\xi}\bar{L}}\bm{Q}_1}\bigg)\times  \\
     & \langle X^{\dagger \xi}_{L, \bm{Q}+\bm{q}} X^{\dagger\tilde{\xi}}_{\tilde{L}, \bm{Q}_1-\bm{q}}X^{\bar{\xi}}_{\bar{L}, \bm{Q}_1}\rangle\nonumber  
 \end{align}
 from which we may construct an exchange matrix element such that 
 \begin{equation}
     \frac{d}{dt}\langle X^{\dagger \xi'}_{L', \bm{Q}}\rangle |_{\mathrm{exch.}}=\frac{i}{\hbar}\sum_{\substack{\bm{q}, \bm{Q}_1\\ \xi, \tilde{\xi}, \bar{\xi}\\ L, \tilde{L}, \bar{L}}} E^{\xi\tilde{\xi}\bar{\xi}\xi'}_{L, \tilde{L},\bar{L}, L', \bm{Q}, \bm{Q}_1, \bm{q} }\langle X^{\dagger \xi}_{L, \bm{Q}+\bm{q}} X^{\dagger\tilde{\xi}}_{\tilde{L}, \bm{Q}_1-\bm{q}}X^{\bar{\xi}}_{\bar{L}, \bm{Q}_1}\rangle \ , 
 \end{equation}
 where the exchange part of the exciton-exciton interaction reads
 \begin{align}
\label{exchangeint}
 \begin{split}
     E^{\xi_1\xi_2\xi_3\xi_4}_{L_1, L_2, L_3, L_4, \bm{Q}_1, \bm{Q}_2, \bm{q}}&=\frac{1}{2}\sum_{\substack{\bm{k},\bm{k}'}} \bigg((V^{c_{l_{1,e}}v_{l_{4,h}}}_{\bm{k}-\bm{k}'} \varphi^{\xi_4 L_4}_{\bm{k}'-\alpha^{\xi_4 L_4}\bm{Q}_1-\bm{q}}-V^{c_{l_{1,e}}c_{l_{4,e}}}_{\bm{k}-\bm{k}'} \varphi^{\xi_4 L_4}_{\bm{k}-\alpha^{\xi_4 L_4}\bm{Q}_1-\bm{q}})\times \\
    &\delta^{\xi_{1,h}, \xi_{4,h}}_{l_{1,h}, l_{4,h}}\delta^{\xi_{2,e}, \xi_{4,e}}_{l_{2,e}, l_{4,e}}\delta^{\xi_{3,e}, \xi_{1,e}}_{l_{3,e}, l_{1,e}}\delta^{\xi_{3,h}, \xi_{2,h}}_{l_{3,h}, l_{2,h}}\varphi^{*\xi_1 L_1}_{\bm{k}-\alpha^{\xi_1 L_1}(\bm{Q}_1+\bm{q})}\varphi^{*\xi_2L_2}_{\bm{k}'-\beta^{\xi_2L_2}\bm{q}-\alpha^{\xi_2L_2}\bm{Q}_2}\varphi^{\xi_3L_3}_{\bm{k}'-\alpha^{\xi_3L_3}\bm{Q}_2}  \\ 
    & +(V^{c_{l_{4,e}}v_{{l_{1,h}}}}_{\bm{k}-\bm{k}'} \varphi^{\xi_4 L_4}_{\bm{k}'+\beta^{\xi_4 L_4}\bm{Q}_1+\bm{q}}-V^{v_{l_{4,h}}v_{l_{1,h}}}_{\bm{k}-\bm{k}'} \varphi^{\xi_4L_4}_{\bm{k}+\beta^{\xi_4 L_4}\bm{Q}_1+\bm{q}})\times\\
    &\delta^{{\xi}_{1,e}, \xi_{4,e}}_{l_{1,e}, l_{4,e}}\delta^{\xi_{2,h}, \xi_{4,h}}_{l_{2,h}, l_{4,h}}\delta^{\xi_{3,e},\xi_{2,e}}_{l_{3,e}, l_{2,e}}\delta^{\xi_{3,h}, \xi_{1,h}}_{l_{3,h}, l_{1,h}}\varphi^{*{\xi_1}{L_1}}_{\bm{k}+\beta^{{\xi_1}{L_1}}(\bm{Q}_1+\bm{q})}\varphi^{*\xi_2 L_2}_{\bm{k}'+\alpha^{\xi_2 L_2}\bm{q}+\beta^{\xi_2L_2}\bm{Q}_2}\varphi^{\xi_3L_3}_{\bm{k}'+\beta^{\xi_3L_3}\bm{Q}_2}\bigg) \ . 
    \end{split}
 \end{align}
 Here, we note that the first term corresponds to hole-hole exchange within the excitons and the second term corresponds to electron-electron exchange. In particular, the Kronecker deltas imply that fermionic exchange of individual charge constituents is only allowed if charges of the same species reside in the same layer and valley. Moreover, the exchange interaction is generally dependent on both centre-of-mass momenta $\bm{Q}_1, \bm{Q}_2$ as well as the relative momentum $\bm{q}$. In the long wavelength limit ($q, Q_1, Q_2\ll a_B^{-1}$, $a_B$ being the exciton Bohr radius) the exchange interaction is non-vanishing for both intra- and interlayer exciton species, and it is the dominating contribution to the exciton-exciton interaction for intralayer excitons \cite{ciuti1998role, schindler2008analysis, tassone1999exciton}. However, we remark that the resulting density-dependent energy renormalizations due to intralayer exchange interactions are negligible (see Supplementary Section IV). Exchange interactions are therefore not considered in the main manuscript, but included here only for the sake of completeness.
 Hence, we obtain the exchange part of the exciton-exciton Hamiltonian
\begin{equation}
    H_{x-x}|_{\mathrm{exch.}}=\frac{1}{2}\sum_{\substack{\bm{Q}_1,\bm{Q}_2, \bm{q}\\ \xi_1...\xi_4\\ L_1...L_4}} E^{\xi_1\xi_2\xi_3\xi_4}_{L_1, L_2, L_3, L_4, \bm{Q}_1, \bm{Q}_2, \bm{q}}X^{\dagger \xi_1}_{L_1, \bm{Q}_1+\bm{q}}X^{\dagger \xi_2}_{L_2, \bm{Q}_2-\bm{q}}X^{\xi_3}_{L_3, \bm{Q}_2}X^{ \xi_4}_{L_4, \bm{Q}_1} \ . 
\end{equation}
We now have access to the multilayer exciton-exciton interaction involving both intra- and interlayer excitons. However, generally, excitons are hybridized between the layers due to electron/hole tunneling. To include the effect of hybridization, we transform the excitonic operators to the hybrid basis, cf. Eq. \eqref{trans}. The exciton-exciton Hamiltonian then transforms into a hybrid Hamiltonian
\begin{equation}
    \tilde{H}_{x-x}=\frac{1}{2}\sum_{\substack{\eta_1...\eta_4 \\ \xi_1...\xi_4\\ \bm{Q}_1, \bm{Q}_2, \bm{q}}} \tilde{W}^{\xi_1\xi_2\xi_3\xi_4}_{\eta_1, \eta_2, \eta_3, \eta_4, \bm{Q}_1, \bm{Q}_2, \bm{q}}Y^{\dagger \xi_1}_{\eta_1, \bm{Q}_1+\bm{q}}Y^{\dagger \xi_2}_{\eta_2, \bm{Q}_2-\bm{q}}Y^{\xi_3}_{\eta_3, \bm{Q}_2}Y^{ \xi_4}_{\eta_4, \bm{Q}_1} \ , 
    \label{hybridhamilton}
\end{equation}
where the hybrid exciton-exciton interaction contains of a direct part and an exchange part according to
\begin{equation}
    \tilde{D}^{\xi_1\xi_2}_{\eta_1,\eta_2, \eta_3, \eta_4, \bm{Q}_1, \bm{Q}_2, \bm{q}}=\sum_{L_1, L_2}D^{\xi_1\xi_2}_{L_1, L_2, \bm{q}}C^{*\xi_1\eta_1}_{L_1, \bm{Q}_1+\bm{q}}C^{\xi_1\eta_4}_{L_1, \bm{Q}_1}C^{*\xi_2\eta_2}_{L_2, \bm{Q}_2-\bm{q}}C^{\xi_2\eta_3}_{L_2, \bm{Q}_2} \ ,
    \label{directs}
\end{equation}
and 
\begin{equation}
    \tilde{E}^{\xi_1\xi_2\xi_3\xi_4}_{\eta_1,\eta_2, \eta_3, \eta_4, \bm{Q}_1, \bm{Q}_2, \bm{q}}=\sum_{L_1, L_2, L_3, L_4}E^{\xi_1\xi_2\xi_3\xi_4}_{L_1, L_2,L_3, L_4,\bm{Q}_1, \bm{Q}_2,   \bm{q}}C^{*\xi_1\eta_1}_{L_1, \bm{Q}_1+\bm{q}}C^{\xi_4\eta_4}_{L_4, \bm{Q}_1}C^{*\xi_2\eta_2}_{L_2, \bm{Q}_2-\bm{q}}C^{\xi_3\eta_3}_{L_3, \bm{Q}_2} \ , 
\end{equation}
with the unhybrised direct ($D$) and exchange ($E$) matrix elements defined in Eq. \eqref{dirint} and \eqref{exchangeint}, respectively. 
Having derived the most general form of the hybrid exciton-exciton interaction, we now remark on the considered case of untwisted homobilayers. In this case, it holds that the mixing coefficients are approximately constant in momentum, such that $C^{\xi\eta}_{L, \bm{Q}}\approx{C^{\xi\eta}_{L}} $ \cite{hagel2021exciton}. Consequently, the direct hybrid exciton-exciton interaction depends only on the relative momentum $\bm{q}$. In the main manuscript, we only consider the lowest-lying hybrid exciton states for each valley configuration and therefore the indices $\eta_i$, $i=1...4$ are omitted therein. Moreover, note that the intra- and interlayer mixing coefficients enter the hybrid exciton-exciton interaction strengths. This provides an intriguing way of tuning the interaction strength with externally applied electric fields. 
\section{Dipole-dipole interaction}
In here, we show that the real-space representation of the direct interlayer exciton-exciton interaction (Eq. \eqref{dirint}, with $L=L'=IX$ and $\xi=\xi'$) can be interpreted as a classical dipole-dipole interaction at large distances. By considering the two TMD layers forming the homobilayer as two infinitely thin slabs separated by the distance $d$ and approximating the dielectric environment as homogenous, with a single effective dielectric constant $\epsilon_{\mathrm{BL}}$, we find an analytic expression for the direct exciton-exciton interaction. Within these approximations, the intra ($X$)- and interlayer ($IX$) Coulomb interactions read \cite{ovesen2019interlayer}
\begin{equation}
    V^{X}_{\bm{q}}=\frac{e_0^2}{2\epsilon_0 A |\bm{q}| \epsilon_{\mathrm{BL}}} \ , V^{IX}_{\bm{q}}=\frac{e_0^2}{2\epsilon_0 A |\bm{q}| \epsilon_{\mathrm{BL}}(1+\mathrm{tanh}(|\bm{q}|d))} \ . 
\end{equation}
Now, substituting these simplified expressions into the direct exciton-exciton interaction and setting the excitonic form factors $F\approx{1}$, we find that the direct interlayer exciton-exciton interaction reduces to  
\begin{equation}
    D_{IX, IX, \bm{q}}\approx{\frac{e_0^2}{2\epsilon_0 A |\bm{q}| \epsilon_{\mathrm{BL}}} (1-\frac{1}{1+\tanh(|\bm{q}|d)})} \ . 
\end{equation}
The real-space representation of the interaction is obtained by taking the Fourier-transform: 
\begin{equation}
    D_{IX, IX}(\bm{r})=\frac{e_0^2}{4\pi \epsilon_0 \epsilon_{\mathrm{BL}}}(\frac{1}{|\bm{r}|}-\frac{1}{d(\sqrt{4+|\bm{r}|^2/d^2)}}) \ . 
\end{equation}
Finally, we are interested in the asymptotic behavior of the interaction and therefore let $r\gg d$. In this limit, we find 
\begin{equation}
    D_{IX, IX}(\bm{r})=\frac{d^2e_0^2}{2\pi\epsilon_0\epsilon_{\mathrm{BL}}}\frac{1}{|\bm{r}|^3}+\mathcal{O}(|\bm{r}|^{-5}) \ , 
\end{equation}
which is precisely a classical dipole-dipole interaction. Now, as the interlayer components of the mixing coefficients are approximately constant in momentum (cf. Supplementary Section I) for untwisted TMD homobilayers, we note that also the corresponding real-space hybrid exciton-exciton interaction obeys the asymptotic $1/r^3$ behavior for large $r$. In Fig. S1, we show the direct hybrid exciton-exciton interaction strength between K$\Lambda$ excitons in hBN-encapsulated WSe$_2$ homobilayers as a function of distance, $r$, including the dominant interlayer contributions to the interaction (solid yellow curve). Unlike in the main manuscript, we here plot the logarithm of the interaction strength over large distances. Importantly, we find that the interaction scales as $1/r^3$ (dashed black curve) at distances $r\gtrsim15$ nm. This confirms the dipole-dipole-like character of the hybrid exciton-exciton interaction at large distances. 
\begin{figure}[h!]
    \centering
    \includegraphics{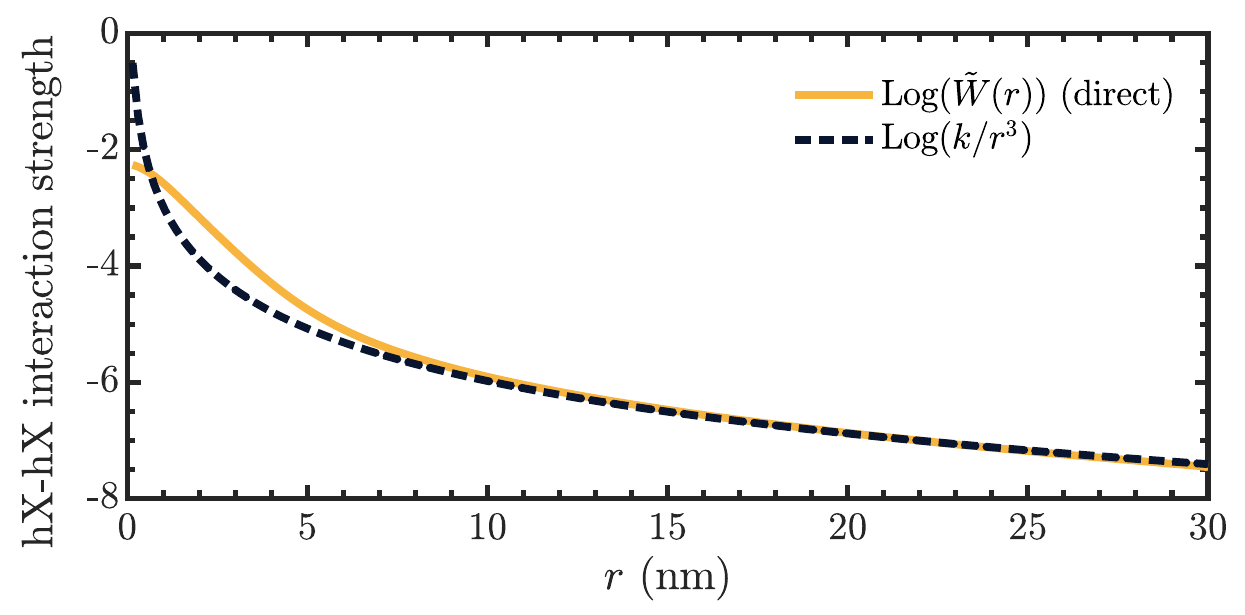}
    \caption{Real-space representation of the hybrid exciton-exciton interaction for K$\Lambda$ excitons. Note that we here have taken the logarithm of the interaction strength. }
    \label{fig:my_label}
\end{figure}
\section{Density-dependent energy renormalizations for hybrid excitons}
\label{energyrensec}
 Having access to the microscopic hybrid exciton-exciton Hamiltonian (Eq. \eqref{hybridhamilton} and Supplementary Section \ref{interactionsec}) implies that we are now also able to investigate density-dependent energy renormalizations of hybrid excitons. These energy renormalizations can be derived from the corresponding Heisenberg equation of motion for the (hybrid) polarisation $\langle Y^{\dagger}_{\zeta, \bm{Q}}\rangle $ reading 
 \begin{align}
     \frac{d}{dt}\langle Y^{\dagger}_{\zeta, \bm{Q}}\rangle=\frac{i}{\hbar}\sum_{\bm{Q}_1, \bm{q}, \zeta_1, \zeta_2, \zeta_3}\tilde{W}^{\zeta_1 \zeta_2 \zeta_3 \zeta}_{\bm{Q}, \bm{Q}_1, \bm{q}}\langle Y^{\dagger}_{\zeta_1, \bm{Q}+\bm{q}}Y^{\dagger}_{\zeta_2, \bm{Q}_1-\bm{q}}Y_{\zeta_3, \bm{Q}_1}\rangle \ , 
 \end{align}
 where we introduced the compound index $\zeta=(\xi, \eta)$ including the valley index $\xi$ and the hybrid exciton index $\eta$. Now, we consider the equation above on a Hartree-Fock level, i.e. we expand the appearing bosonic three-operator expectation value into single-particle expectation values and neglect two-particle correlations. We also make use of the random phase approximation (RPA) \cite{kira2006many} such that the equation above becomes 
  \begin{align}
     \frac{d}{dt}\langle Y^{\dagger}_{\zeta, \bm{Q}}\rangle\approx{\frac{i}{\hbar}\sum_{\bm{q}, \zeta_1}(\tilde{W}^{\zeta \zeta_1 \zeta_1 \zeta}_{\bm{Q}, \bm{q}, 0}+\tilde{W}^{\zeta_1 \zeta \zeta_1 \zeta}_{\bm{Q}, \bm{q}, \bm{q}-\bm{Q}})N^{\zeta_1}_{\bm{q}} \langle Y^{\dagger}_{\zeta, \bm{Q}}\rangle } \ , 
 \end{align}
 where we defined the hybrid exciton occupation $N^{\zeta}_{\bm{Q}}\equiv \langle Y^{\dagger}_{\zeta, \bm{Q}}Y_{\zeta, \bm{Q}}\rangle$. Here we also approximated $\langle Y^{\dagger}_{\zeta, \bm{Q}}Y_{\zeta', \bm{Q}'}\rangle \approx{\delta^{\zeta, \zeta'}_{\bm{Q}, \bm{Q}'} N^{\zeta}_{\bm{Q}}}$ making use of the RPA. Note that the energy renormalization consists of two terms, the first term being a direct term and the second being an exchange term, with the latter reflecting exciton exchange \cite{ciuti1998role}. In contrast to the exchange interaction $\tilde{E}$ which includes exchange of individual carriers, the exciton exchange takes into account the exchange of individual excitons. Furthermore, we consider low temperatures in this work such that the exciton distribution $N_{\bm{q}}$ is strongly peaked around $\bm{q}=0$ and assume the centre-of-mass momentum $|\bm{Q}|\ll a_B^{-1}$, where $a_B$ is the exciton Bohr radius. This reduces the equation above to $\frac{d}{dt}\langle Y^{\dagger}_{\zeta, \bm{Q}}\rangle \approx{\frac{i}{\hbar}\delta E^{\zeta} \langle Y^{\dagger}_{\zeta, \bm{Q}}\rangle } $ introducing the energy renormalization
 \begin{equation}
     \delta E^{\zeta}=A \sum_{\zeta_1} (\tilde{W}^{\zeta_1 \zeta \zeta_1 \zeta}_{0, 0, 0}+\tilde{W}^{\zeta \zeta_1 \zeta_1 \zeta}_{0, 0, 0})n_x^{\zeta_1} \ , 
 \end{equation}
where $n^{\zeta_1}_x\equiv \frac{1}{A}\sum_{\bm{Q}}N^{\zeta_1}_{\bm{Q}}$ with $A$ being the crystal area (cancelling out with the area $A$ in the electronic Coulomb matrix elements). Finally, we restrict ourselves to the energetically lowest hybrid exciton states in this work such that the compound index $\zeta$ reduces to the valley index $\xi$. Generally, the energy renormalization of a hybrid exciton $\xi$ is obtained by taking into account the interactions between all the different exciton species. By assuming that $n_x^{\xi}\propto n_x$, i.e. assuming a thermalized Boltzmann distribution for the hybrid excitons, where $n_x=\sum_{\xi} n^{\xi}_x$ is the total exciton density, the energy renormalization of a single exciton species can be quantified by a valley-specific effective dipole length $d^{\xi}$ obtained from 
\begin{equation}
    \delta E^{\xi}=A \sum_{\xi_1} (\tilde{W}^{\xi_1 \xi \xi_1 \xi}_{0, 0, 0}+\tilde{W}^{\xi \xi_1 \xi_1 \xi}_{0, 0, 0})n_x^{\xi_1}\equiv \frac{d^{\xi}e_0^2}{\epsilon_0 \epsilon^{\perp}_{\mathrm{TMD}}}n_x \ . 
    \label{generalenergyren}
\end{equation}
In the evaluation of valley-specific dipole lengths we here only include the direct (dipole-dipole) contributions to the hybrid exciton-exciton interaction, cf. Eq. \eqref{hybridhamilton}. This is done as interlayer exchange interactions are seen to provide a small quantitative correction to the dipole-dipole interaction \cite{erkensten2022microscopic}. Moreover, although intralayer exchange interactions taking into account individual exchange of carriers are dominant in TMD monolayers \cite{PhysRevB.96.115409, erkensten2021exciton}, they have negligible impact on the energy renormalizations, as their contributions are largely cancelled out against contributions due to higher-order correlations such as biexcitons \cite{katsch2019theory}. This goes beyond the scope of the Hartree-Fock theory presented in this work.

\begin{figure}[h!]
     \centering
     \includegraphics[width=\columnwidth]{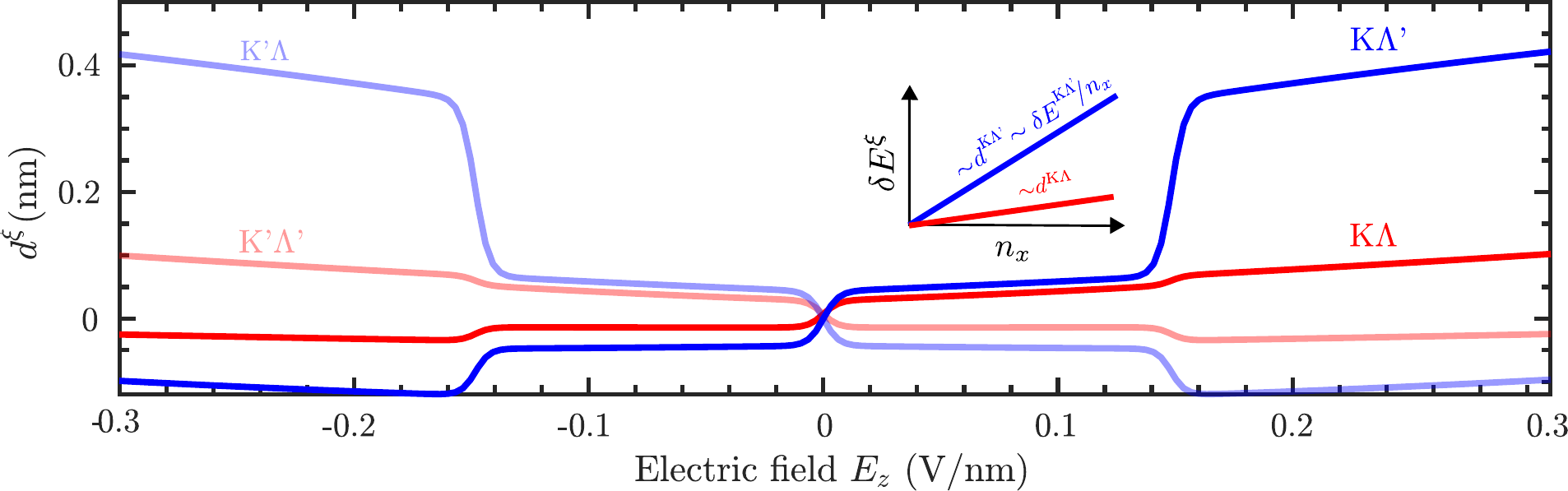}
     \caption{Valley-specific effective dipole length in naturally stacked hBN-encapsulated WSe$_2$ homobilayers. The valley-specific dipole length $d^{\xi}$ determines the corresponding energy renormalization $\delta E^{\xi}\sim d^{\xi}n_x$ for an individual exciton species $\xi$ (cf. inset).}
     \label{valleydipole}
 \end{figure}

In Fig. \ref{valleydipole} we illustrate the valley-specific effective dipole length as obtained from Eq. \eqref{generalenergyren} for the case of naturally stacked hBN-encapsulated WSe$_2$ homobilayers as a function of electric field, $E_z$ at low temperatures, $T=10$ K. 
 Intriguingly, we find a drastic increase in the dipole length at around $E_z\pm 0.15$ V/nm. This reflects the transition from a mostly intralayer K$\Lambda$ (K'$\Lambda$') state to a mostly interlayer K$\Lambda$' (K'$\Lambda$) state under the application of a positive (negative) electric field. At vanishing electric fields, K$\Lambda$ and K'$\Lambda$' excitons coexist. These excitons independently interact via weak repulsive dipole-dipole interactions and mutually interact with each other via attractive dipolar interactions as they exhibit opposite dipole orientations, giving rise to negligible effective dipole lengths. At the largest considered electric fields, $E_z=0.3$ V/nm, only K$\Lambda$' excitons are relevant and the impact of other excitons is negligible due to the large energy separations between exciton states (cf. Table \ref{energytable}). These excitons are mostly interlayer-like in nature ($|C_{IX}|^2=0.8$, cf.  Table \ref{energytable}) and interact strongly via dipole-dipole repulsion. Note that the large effective dipole moment of K$\Lambda$' excitons directly translates into large energy renormalizations (Eq. \eqref{generalenergyren}) and give rise to sizable blue-shifts of phonon sidebands with exciton density, as schematically illustrated by the inset in Fig. \ref{valleydipole}. 
\newpage